\documentclass[fleqn,usenatbib]{mnras}

\usepackage{newtxtext,newtxmath}
\usepackage{color}

\usepackage[T1]{fontenc}

\DeclareRobustCommand{\VAN}[3]{#2}
\let\VANthebibliography\thebibliography
\def\thebibliography{\DeclareRobustCommand{\VAN}[3]{##3}\VANthebibliography}

\usepackage{graphicx}	
\usepackage{amsmath}	
\usepackage{float}      
\usepackage{widetext} 
\usepackage{ulem}

\title[Filament spin detection with kSZ]{Measuring cosmic filament spin with the kinetic Sunyaev-Zel'dovich effect}

\author[Y. Zheng et al.]{
Yi Zheng,$^{1,2,3}$\thanks{E-mail: zhengyi27@mail.sysu.edu.cn}
Yan-Chuan Cai,$^{4}$\thanks{E-mail: cai@roe.ac.uk}
Weishan Zhu,$^{1,2}$\thanks{E-mail: zhuwshan5@mail.sysu.edu.cn}
Mark Neyrinck,$^{5,6,7}$
Peng Wang,$^{8,9}$
Shaohong Li$^{1}$
\\
$^{1}$School of Physics and Astronomy, Sun Yat-sen University, 2 Daxue Road, Tangjia, Zhuhai, 519082, China\\
$^{2}$CSST Science Center for the Guangdong-Hong kong-Macau Greater Bay Area, SYSU\\
$^{3}$Key Laboratory for Particle Astrophysics and Cosmology (MOE)/Shanghai Key Laboratory for Particle Physics and Cosmology, China\\
$^{4}$Scottish Universities Physics Alliance, Institute for Astronomy, University of Edinburgh, Blackford Hill, Scotland, UK\\
$^{5}$Ikerbasque, the Basque Foundation for Science, 48009, Bilbao, Spain\\
$^{6}$Department of Physics, University of the Basque Country UPV/EHU, 48080, Bilbao, Spain\\
$^{7}$Donostia International Physics Center, Paseo Manuel Lardizabal 3, 20018 San Sebasti\'an, Spain\\
$^{8}$Shanghai Astronomical Observatory, CAS, Nandan Road 80, Shanghai 200030, China\\
$^{9}$Key Laboratory for Research in Galaxies and Cosmology, SHAO, CAS, Nandan Road 80, Shanghai 200030, China}

\date{Accepted XXX. Received YYY; in original form ZZZ}

\pubyear{2022}

\newcommand{\beq}{\begin{equation}}
\newcommand{\eeq}{\end{equation}}
\newcommand{\bea}{\begin{eqnarray}}
\newcommand{\eea}{\end{eqnarray}}
\newcommand{\bi}{\begin{itemize}}
\newcommand{\ei}{\end{itemize}}
\newcommand{\bfi}{\begin{figure}[!t]
\epsfxsize=9cm
\epsffile}
\newcommand{\bfig}{\begin{figure*}[!t]
\center{}
\epsfxsize=15cm
\epsffile}
\newcommand{\efi}{\end{figure}}
\newcommand{\efig}{\end{figure*}}
\newcommand{\no}{\nonumber}
\newcommand{\mpch}{{\rm Mpc}/h}

\newcommand{\bfv}{{\bf v}}
\newcommand{\tauT}{\tau_{\rm T}}

\newcommand{\muk}{\mu \rm K}

\newcommand{\ksz}{{\rm kSZ}}

\definecolor{ForestGreen}{rgb}{0.3,0.7,0.3}

\begin{document}
\label{firstpage}
\pagerange{\pageref{firstpage}--\pageref{lastpage}}
\maketitle


\begin{abstract}
The spin of intergalactic ﬁlaments has been predicted from simulations, and supported by tentative evidence from redshift-space ﬁlament shapes in a galaxy redshift survey: generally, a ﬁlament is redshifted on one side of its axis, and blueshifted on the other. Here, we investigate whether ﬁlament spins could have a measurable kinetic Sunyaev-Zel’dovich (kSZ) signal, from CMB photons being scattered by moving ionised gas; this pure velocity information is complementary to ﬁlament redshift-space shapes. We propose to measure the kSZ dipole by combining galaxy redshift surveys with CMB experiments. We base our S/N analyses ﬁrst on an existing ﬁlament catalogue, and its combination with Planck data. We then investigate the detectability of the kSZ dipole using the combination of DESI or SKA-2 with next-stage CMB experiments. We ﬁnd that the gas halos of ﬁlament galaxies co-rotating with ﬁlaments induce a stronger kSZ dipole signal than that from the diﬀuse ﬁlamentary gas, but both signals seem too small to be detected in near-term surveys such as DESI+future CMB experiments. But the combination of SKA-2 with future CMB experiments could give a more than 10$\sigma$ detection. The gain comes mainly from an increased area overlap and an increased number of ﬁlaments, but also the low noise and high resolution in future CMB experiments are important to capture signals from ﬁlaments small on the sky. Successful detection of the signals may help to ﬁnd the gravitomagnetic eﬀect in large-scale structure and advance our understanding of baryons in the cosmic web.
\end{abstract}


\section{Introduction}
\label{sec:Intro}

Spin is a general property of celestial bodies in our Universe. It is related to large-scale structure formation and encodes information about cosmology and astrophysics. Understanding the spin of matter on large scales have several implications: (1) Spin is part of the cosmic curl velocity field generated by the nonlinear structure formation from the primordial fluctuations. Understanding the spin properties of structures within different cosmic-web environments helps us tackle the late-time cosmic structure formation at nonlinear scales  \citep{Pichon1999,Zheng13,Libeskind2014,Zhu2015,Zhu2017,Sheng2022} or exploit information from the early universe \citep{YuHR2020,Motloch2021,Motloch2022}; (2) The spin of matter should in principle induce vector potentials. This will give rise to an additional gravitational lensing signal, known as the gravitomagnetic effect \citep[e.g.][]{Sereno2002,Sereno2003,Schafer2006,Cristian2021,Cristian2022}. It is a prediction from general relativity but has not been detected at cosmological scales \citep{Tang2021}. (3) The spin of filaments may have implications for the formation of galaxies. The spin directions of certain galaxies around filaments are known to correlated with the directions of the filaments, and so the spin of filaments may add sophistication to this physical picture (e.g.  \citealt{WangP2017,WangP2018,WangP2018b}). Another reason to study filament spin is if it substantially affects the spin of galaxies inside them.

For several years, filaments have been suspected to be `swirling, rotating environments' that can impart spin to halos \citep{codis/etal:2012,laigle/etal:2015}. It has also been qualitatively predicted to spin coherently themselves \citep{neyrinck/2016,Neyrinck2020}. This is perhaps not surprising; generally, things in the Universe have some spin unless there is a reason they should not. Larger cosmic-web elements, walls and voids, are not thought to spin coherently and substantially in a universe without substantial initial vorticity, because they expand rather than contract along any conceivable spin axis.

Quite recently, this prediction of filament spin has been quantitatively confirmed in simulations and observations. By stacking the angular momentum of the intergalactic filaments along the filament spine from $N$-body simulations, \citep[][X21 hereafter]{Xia2021} found that they typically spin coherently at tens or hundreds of km/s. Essentially simultaneously, there was also a tentative observational detection \citep[][W21 hereafter]{WangP2021} of filament spin, for SDSS (Sloan Digital Sky Survey) galaxies in filaments. By analysing the average redshifts of galaxies around stacked filaments that are nearly perpendicular to the line of sight, W21 found a $\sim$3$\sigma$ redshift dipole for galaxies on either side of filament axes. One explanation for the dipole is that those filaments spin, and they claim a rotation curve with a peak of 80$\sim$100 km/s at $\sim$1 Mpc to the filament spine.

But further theoretical and observational work is necessary to clarify the picture.  The results in W21 are consistent with spin, and that seems to be the most likely explanation, but other possibilities are conceivable. Their key observation is that in redshift space, filaments nearly in the plane of the sky are typically substantially tilted along the line of sight. Since galaxy positions and velocities are typically unknown independently outside the very local universe, this redshift-space signal is perhaps the best way to test the hypothesis of rotating filaments. But instead of rotation, the signal could in principle be produced by a shearing motion. Or, it may not come from redshift-space distortions at all; if a filament in the plane of the sky is elliptical in cross section, with a major axis diagonal between the plane of the sky and the line of sight, this could mimic the redshift dipole. But a good argument that the measurement genuinely measured spin is that the rotation velocities in W21 and X21, coming from independent analyses analyses using quite different techniques, are remarkably similar.

Here, we concentrate on another observable: a kinetic Sunyaev-Zel'dovich (kSZ) \citep{Sunyaev1980} filament-rotation signal would come purely from the velocity field, removing the ambiguity in the W21 measurement that filaments might be tilted along the line of sight even in real space.

When CMB photons travel through a filament, they are inverse Compton scattered off by free electrons {moving there with respect to the CMB rest frame}. As a result, the temperature of the CMB photons will be slightly {shifted}. The CMB temperature fluctuation induced by the kSZ effect is
\beq
\label{eq:kSZ_basic}
\delta T_{\rm kSZ}(\hat{n}) = -T_0\int dl\sigma_{\rm T} n_{\rm e} \left(\frac{{\bfv}_{\rm e}\cdot \hat{n}}{c}\right)  \,,
\eeq
where $T_0\simeq 2.7255\rm K$ is the averaged CMB temperature, $\sigma_{\rm T}$ is the Thomson-scattering cross-section, $c$ is the speed of light, $\hat{n}$ is the unit vector along the line of sight (LOS), $n_{\rm e}$ is the physical free electron number density, $\bfv_{\rm e}$ is the physical peculiar velocity of free electrons, defined to be positive for those recessional objects, and the integration $\int dl$ is along the LOS given by $\hat{n}$ \citep{Sunyaev1980}.

On average, there are equal probabilities of electrons moving away and towards the CMB, so the expected line-of-sight integral is zero. Yet the rotational component of the momentum field will survive the line-of-sight integration (e.g.  \citealt{Zhang2004,Shao11b}). In~\cite{Matilla2020}, it was proposed that the spinning gaseous galactic halos can be detected via a rotational kSZ effect. Likewise, a spinning filament with the direction of its angular momentum perpendicular to the line of sight will also leave a kSZ temperature dipole imprinted on the CMB, with its amplitude being proportional to the electron density and the rotational velocity. This kSZ dipole should be a relatively unique feature associated with the spin of filaments. Its detection would therefore serve as an independent evidence for the spin of filaments. Meanwhile, a detection of this dipolar kSZ signal from filaments will also provide information about the abundance of free electron in filaments. This will help to constrain the baryon content in filaments.

Measurements from the cosmic microwave background(CMB) and primeval abundance of light nuclei indicate that ordinary baryonic matter makes up $\sim 5\%$ of the cosmic energy density (e.g.  \citealt{PLANCK2015,PLANCK2018}). The baryon density at redshift $\sim 3$, derived from the Lyman-$\alpha$ forest, is consistent with measurements derived from CMB and light nuclei \citep{1997ApJ...489....7R, 1997ApJ...490..564W}. However, until recent years, baryons found at low redshift seem to add up only a fraction of the total in the standard $\Lambda$CDM model \citep{2012ApJ...759...23S}. New independent measurements seem to have resolve this issue \citep[e.g.][]{Macquart2020}, but localising where the baryons are reminds challenging \citep{Driver2021}.

For a few decades, there has been theoretical (e.g.~\citealt{1970A&A.....5...84Z, 1996Natur.380..603B}) and observational (e.g.  \citealt{1986ApJ...302L...1D}) work indicating that matter is arranged in a cosmic web on scales larger than galaxies, i.e., made of nodes/clusters, filaments, sheets/walls and voids. Meanwhile, cosmological $N$-body/hydrodynamical simulations have been used to reveal the formation and evolution of cosmic web {in quantitative detail} (e.g.  \citealt{2007A&A...474..315A, Cautun2014, Zhu2017}). {These predict that filaments should} host about one half of the baryonic gas in the universe after $z=2$, and should contain most of the missing baryons (e.g.  \citealt{1999ApJ...514....1C, 2019MNRAS.486.3766M}). 

Yet, it remains a challenge to observe the baryons in cosmic filaments. Cosmological hydrodynamical simulations have predicted that baryons residing in filaments are mainly in the `warm-hot' state, with temperatures $10^5-10^7 \rm{K}$, and with density $\sim 1-100$ times the cosmic mean baryon density (e.g.  \citealt{1999ApJ...514....1C, 2001ApJ...552..473D}). Considering the thermal state of baryons in filaments,  X-ray emission and absorption have been often proposed as tools to detect the gas in filaments. Indeed, in the past two decades, a number of works have reported the detection of baryons in filaments in the X-ray (e.g.  \citealt{2002ApJ...572L.127F,2007ARA&A..45..221B, 2015Natur.528..105E, 2018Natur.558..406N, 2020A&A...643L...2T}), and more recently through the thermal Sunyaev-Zel'dovich (tSZ) effect \citep{2018A&A...609A..49B, 2019A&A...624A..48D, 2019MNRAS.483..223T, 2020A&A...637A..41T}. Despite these remarkable achievements, more investigation is needed to improve the significance of signal and reduce the uncertainty. For instance, information of the gas density and temperature in filaments are needed to interpret the reported tSZ signal in recent studies. Also, the techniques above are only sensitive to the baryons with a certain temperature range. On the contrary, the amplitude of the kSZ signal does not depend on the temperature of the electrons. It therefore serves as a powerful tool of searching for the missing baryons in our universe \citep{Shao11b,Carlos2015,Shao2016,Lim2020,Jonas2021}. A robust detection of the kSZ signal associated with the spin of filaments can help to constrain the amount of baryons in filaments.

In view of its wide application prospect, we aim to develop a technique to measure the kSZ dipole induced by the filament spin using the combination of galaxy redshift surveys with CMB experiments. 
This is the theme of this paper, which is organized as follows. In section \ref{sec:theory} we introduce the method of kSZ dipole detection, present the theoretical framework for calculating the expected dipole signal and the associated detection noise given a combination of galaxy redshift surveys with CMB experiments. In section \ref{sec:ston_estimation}, we use this to estimate the kSZ signal-to-noise ratio (S/N) of current and future survey combinations. Section \ref{sec:conclusion} is the conclusion and discussion.

\section{Theoretical setup}
\label{sec:theory}
\begin{figure}
\begin{center}
\includegraphics[width=\columnwidth]{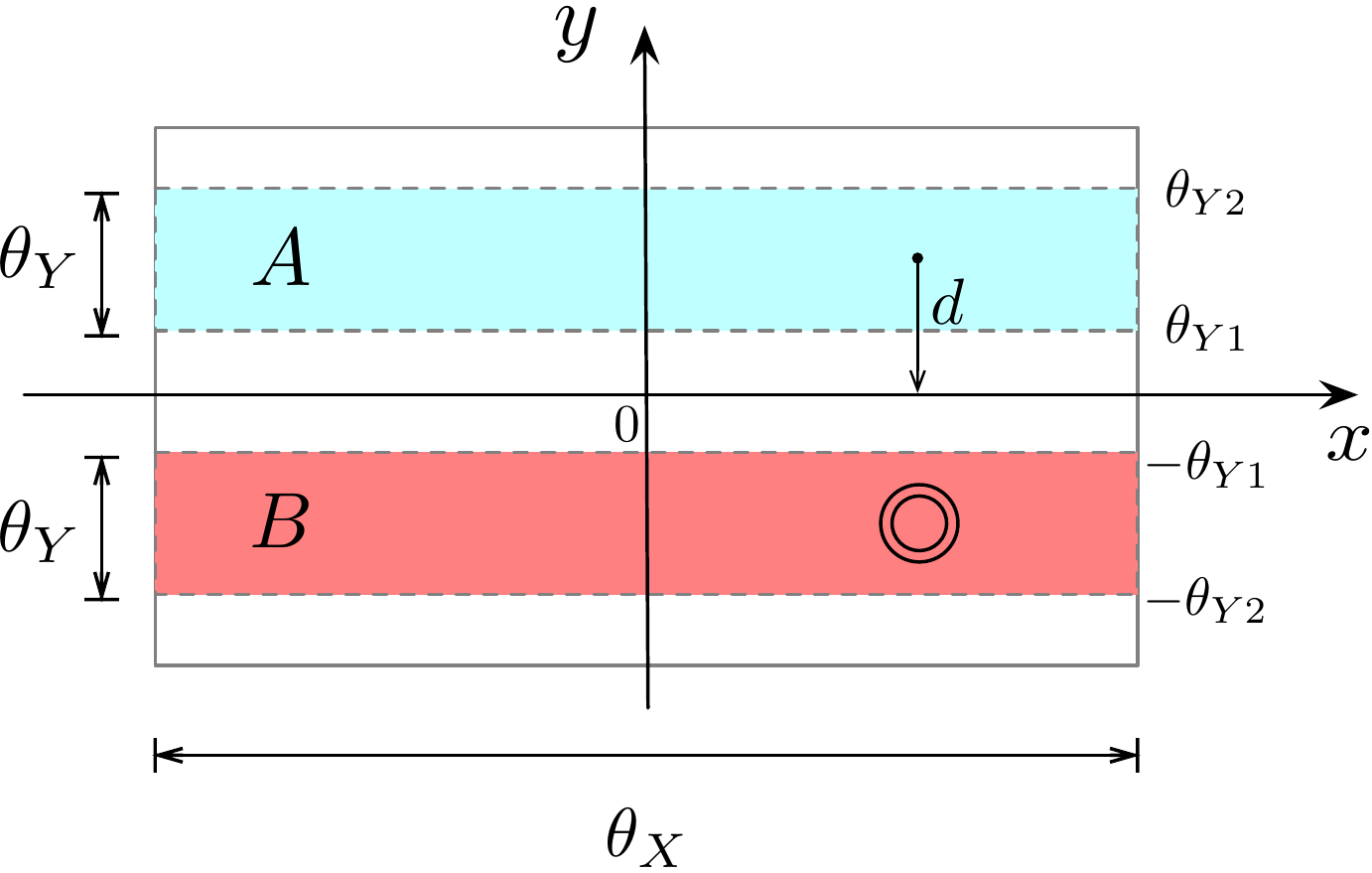}
\end{center}
\caption{Cartoon illustration for the filament projected on the plane of the sky, shown by the gray solid rectangular box. The filament axis/spine is along the $x$ axis.  The rectangular aperture filter $W^{\rm fila}(\vec{\theta})$ is marked by the colored regions A and B. This is the filter we will use to extract the kSZ signal from the spin of diffuse gas around filament. The blue/red side (A/B) represents blueshift/redshift, with matter moving towards/away from us. $\theta_X$ corresponds to the filament length. The two concentric circles represent the compensated filter we will use for extracting the kSZ signal around galaxies co-rotating with the filament.}
\label{fig:filter}
\end{figure}
In this section we outline the theoretical setup for estimating the S/N of the kSZ signals due to the spin of filaments. We will assume that the spin directions of filaments are known from the averaged redshift of galaxies around each filament (W21). Therefore the kSZ dipole is likely to be correlated with the galaxy redshift dipole due to the prior knowledge brought in from the redshift survey. However, this correlation is not complete, since the kSZ probes only the velocity field. Then we can calculate the amplitude of the kSZ dipole, factoring in sample variance and the noise from the CMB. So measuring the kSZ dipole requires the overlap of a galaxy redshift survey with the CMB experiment on the sky.

The major steps are as follows, and illustrated in Fig.~ \ref{fig:filter}:

(1) We identify filaments from a galaxy redshift survey and measure their orientations, which are specified by the viewing angle $\phi$ between the filament axis and the line of sight (LOS).

(2) In each filament, we identify `A' and `B' regions, separated by the filament spine, by specifying their redshift difference $\Delta z_{\rm AB} = \bar{z}_{\rm A} - \bar{z}_{\rm B}$. Here $\bar{z}$ is the averaged redshift of member galaxies within each region. Region-A is defined to have a lower redshift and move towards us, while region-B has a higher redshift and move away from us (Fig. \ref{fig:filter}).

(3) For the $i$th filament, we measure the averaged CMB temperature $T_{{\rm CMB},i}^{A,B}$ of the two regions due to the ionised gas in the filament, and the kSZ signal is measured by $\delta T^{\rm fila}_{{\rm kSZ},i} = \delta T^A_{{\rm kSZ},i} -\delta T^B_{{\rm kSZ},i} = T_{{\rm CMB},i}^A-T_{{\rm CMB},i}^B +\epsilon_i$, where $\epsilon_i$ is the residual noise after the subtraction.
Then we average over all filaments to get $\delta T^{\rm fila}_{\rm kSZ}=\sum_i \delta T^{\rm fila}_{{\rm kSZ},i}/\sum_i$. This signal will be referred to as the `kSZ signal from diffuse gas rotating around filaments' hereafter. 

(4) We also estimate the kSZ signal associated with the gas halo for each galaxy within the region-A and -B, and take the difference of their mean: $\delta T^{\rm fila,gal}_{{\rm kSZ}} = \sum_l \delta T^{A,\rm gal}_{{\rm kSZ},l}/\sum_l-\sum_m \delta T^{B,\rm gal}_{{\rm kSZ},m}/\sum_m$. This signal will be referred to as the `kSZ signal from galaxies rotating around filaments'.

The above steps (3) and (4) illustrate two ways of detecting the kSZ signals due to the spin of filaments, assuming that the galaxies rotates in the same way as the gas does around filaments. The difference is that the `kSZ signal from diffuse gas rotating around filaments' detects all free electrons rotating around the filaments, while the `kSZ signal from galaxies rotating around filaments' mainly comes from the electrons within the gas halos around galaxies associated with the filaments. The mean gas density of filaments is expected to be relatively low, order of 10 $\bar \rho_{\rm baryon}$ (e.g.  \citealt{Zhu2021}, hereafter Z21), while the gas density around galaxies should be much higher and it results in a higher kSZ signal, albeit with a smaller aperture, the noise level for the latter may also be larger. It is the main goal of this study to investigate the S/N's for both scenarios.

\subsection{Filters \& noise}

To model the observed kSZ signal, we break it down into several intermediate steps. First, the physical kSZ signal from equation (\ref{eq:kSZ_basic}) is convolved with the CMB instrumental beam function $B(\vec{\theta},\vec{\theta}')$: 
\beq
\label{eq:T_ksz_obs}
\delta T^{\rm obs}_{\ksz} (\vec{\theta})= \int d^2\theta'B(\vec{\theta}-\vec{\theta}')\delta T_{\ksz}(\vec{\theta}')\,.
\eeq
$\delta T^{\rm obs}_{\ksz}$ is usually overwhelmed by the primordial CMB temperature fluctuations. To reduce this noise, we apply an aperture photometry (AP) filter $W(\vec{\theta},\vec{\theta}')$ at the object's celestial location:
\beq
\label{eq:T_ksz_ap}
\delta T_{\ksz}^{\rm AP}(\vec{\theta})=\int d^2\theta'W(\vec{\theta}-\vec{\theta}')\delta T_{\ksz}^{\rm obs}(\vec{\theta}')\,.
\eeq
The compensated AP filter for filaments $W^{\rm fila}$ consists of two equal areas A\&B, as shown in Fig.~\ref{fig:filter}. The average CMB temperatures of which are subtracted to reduce the primordial CMB temperature fluctuations. For the galaxies, a 2-D compensated top-hat filter $W^{\rm gal}$ is adopted (circles in Fig.~\ref{fig:filter}).

Equations (\ref{eq:T_ksz_obs}) and (\ref{eq:T_ksz_ap}) can be combined in Fourier space as (e.g. \citealt{Sugiyama18})
\beq
\label{eq:T_ksz_AP}
\delta T^{\rm AP}_{\ksz} (\vec{\theta}) = \int \frac{d^2\ell}{(2\pi)^2}e^{i\vec{\ell}\cdot\vec{\theta}}W(\vec{\ell})\delta T_{\ksz}(\vec{\ell})B(\vec{\ell})\,,
\eeq
where $\vec{\ell}$ is the two-dimensional wavevector perpendicular to the LOS, $W(\vec{\ell})$, $\delta T_{\ksz}(\vec{\ell})$ and $B(\vec{\ell})$ are the 2-D Fourier transforms of $W(\vec{\theta})$, $\delta T_{\ksz}(\vec{\theta})$ and $B(\vec{\theta})$ \footnote{The 2-D (inverse) Fourier transform definition we adopt is $f(\vec{\ell})=\int d^2\theta e^{-i\vec{l}\cdot\vec{\theta}}f(\vec{\theta})$, $f(\vec{\theta}) = \int \frac{d^2\ell}{(2\pi)^2}e^{i\vec{\ell}\cdot\vec{\theta}}f(\vec{\ell})$.}. We adopt a Gaussian beam function $B(\vec{\ell}) = e^{-\sigma_{\rm B}^2\ell^2/2}$, in which $\sigma_{\rm B}={\rm FWHM}/\sqrt{8\ln(2)}$, and FWHM is the full width at half-maximum of the beam. It was shown to be accurate for Planck, ACT and SPTpole for most applications (e.g. \citealt{PlanckkSZ16,Calafut2021,Soergel16}). 

We assume that the AP filtered noise fluctuation $\delta T^{\rm AP}_{\rm N}$ is an uncorrelated, Gaussian field that satisfies
\beq
\left<\delta T^{\rm AP}_{\rm N}(\vec{\theta}_i)\delta T^{\rm AP}_{\rm N}(\vec{\theta}_j)\right> = \sigma^2_{\rm N}\delta^D_{ij}\,,
\eeq
and its variance $\sigma^2_{\rm N}$ can then be computed as~\citep{Hernandez2006,Sugiyama18,Zheng20}
\beq
\sigma_{\rm N}^2 = \int \frac{d^2\ell}{(2\pi)^2}[C_\ell B(\vec{\ell})B^\ast(\vec{\ell})+N_\ell]W(\vec{\ell}) W^\ast(\vec{\ell})\,.
\label{eq:noise}
\eeq
Here $C_\ell$ is the lensed CMB angular power spectrum, and $N_\ell$ is the angular power spectrum of the instrumental noise. The intrinsic fluctuations of CMB temperature and the instrumental noise are two major source of noise. They are assumed to be uncorrelated to each other.

\subsection{{kSZ signal from diffuse gas rotating around filaments}}
\label{subsec:theory_filaspin}

\begin{figure}
\centering
\includegraphics[width=0.8\columnwidth]{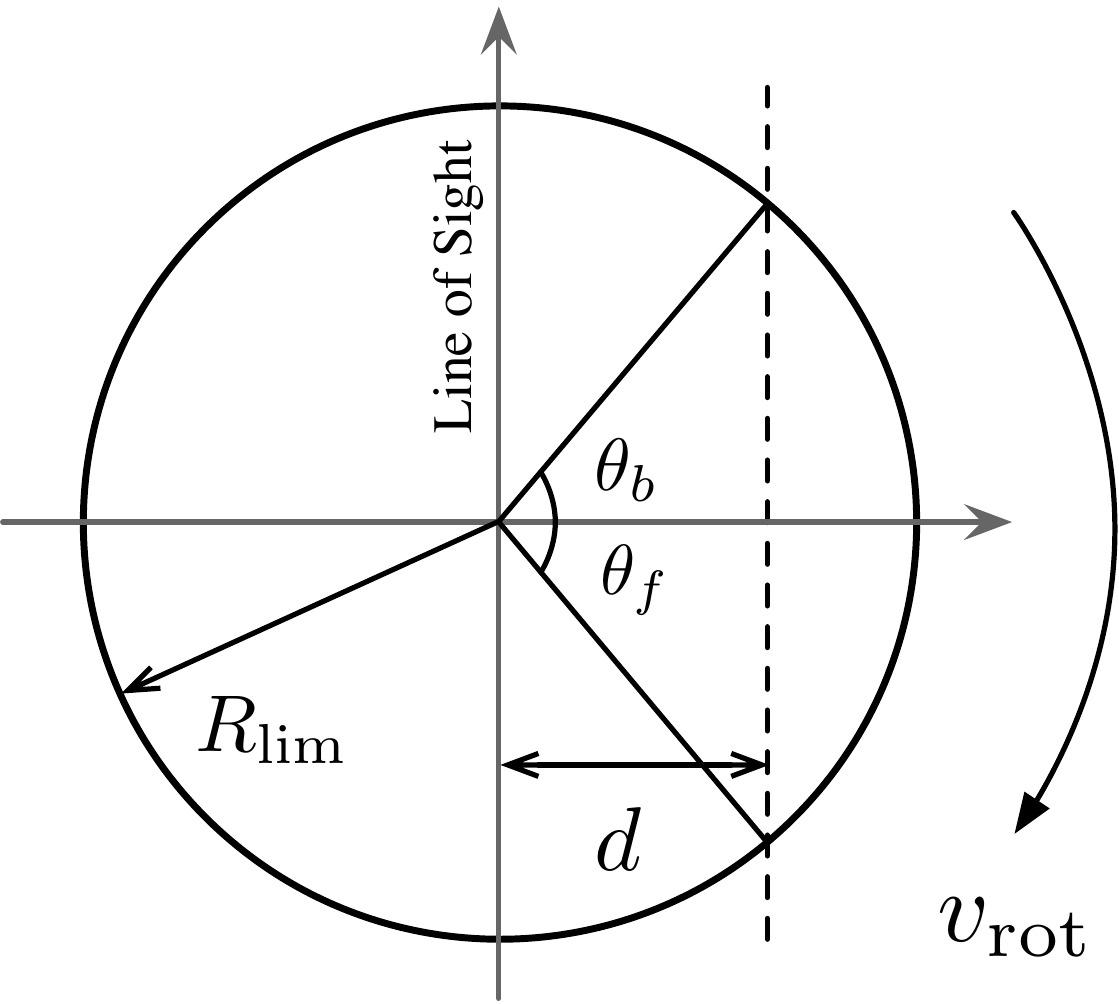}
\caption{Cartoon illustration for the cross section of a filament that lies perpendicular to the LOS and rotates clockwise. The integration of equation (\ref{eq:ksz_profile}) along the LOS is indicated by the dashed line.}
\label{fig:cross_section}
\end{figure}
\begin{figure}
\begin{center}
\includegraphics[width=\columnwidth]{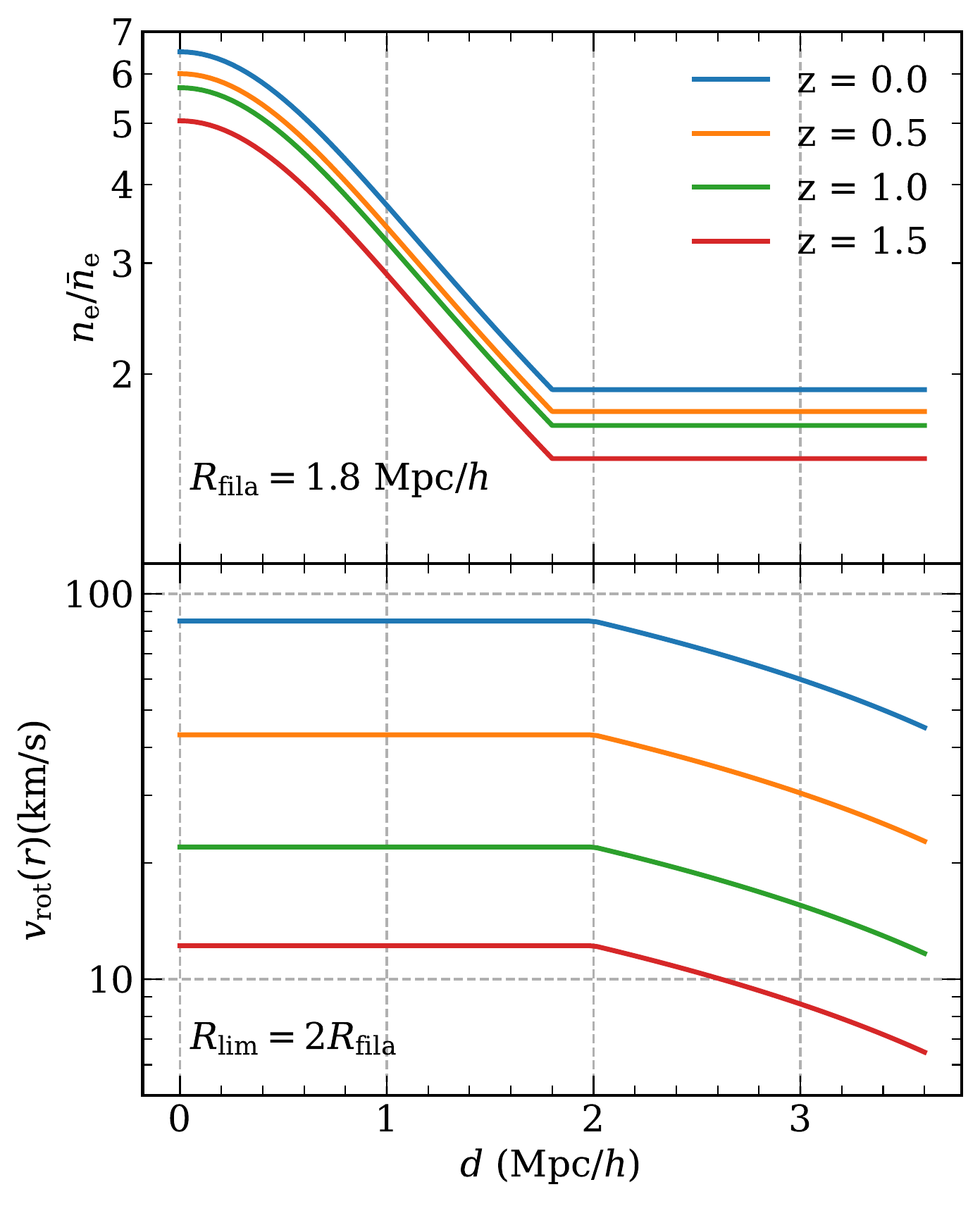}
\end{center}
\caption{Comoving electron number density profiles perpendicular to the spine of filaments (top panel), and the rotation velocity profiles of filaments (bottom panel). Different colours represents different redshifts, as labeled in the legend. $R_{\rm fila}$ gives the boundary of diffuse gas for the filament gas, and $R_{\rm lim}$ indicates the upper bound of our integration for equation (\ref{eq:ksz_profile}).}
\label{fig:prof}
\end{figure}
Next we want to estimate the kSZ signal from diffuse gas rotating around filaments. We start by considering a rotating filament lying perpendicular to the LOS. It is perfectly straight, with ionised gas uniformly distributed along the spine of the filament (Figs. \ref{fig:filter} and \ref{fig:cross_section}). Its kSZ temperature profile perpendicular to the filament spine can be computed from equation (\ref{eq:kSZ_basic}):
\beq
\label{eq:ksz_profile}
\delta T_{\rm kSZ}(d) 
=-\frac{T_0\sigma_{\rm T}}{a^2c}\int_{\theta_f}^{\theta_b}{\rm d}\theta \frac{d}{\cos^2\theta} n_{\rm e}(\frac{d}{\cos\theta})v_{\rm rot}(\frac{d}{\cos\theta})\cos\theta \,,
\eeq
where $a$ is the cosmic scale factor,  $\theta_b = -\theta_f = \cos^{-1}(d/R_{\rm lim})$. $n_{\rm e}(r)$ and $v_{\rm rot}(r)$ are the comoving 3D profiles of the free electron number density and rotation velocity from the filament axis, and $v_{\rm rot}(d/\cos\theta)\cos\theta$ is the component of the rotational velocity projected along the LOS, in which $d$ is the projected comoving distance to the filament spine on the sky. The $a^{-2}$ factor before the integration accounts for the fact that the physical column density of electrons evolves with $(1+z)^2$. The integration on the cross section of a filament is illustrated in Fig.~\ref{fig:cross_section}. {$R_{\rm lim}$ as indicated in the figure is the boundary of integration for the filament, up to which the kSZ signal contribution of baryons in a filament is considered in our evaluation.

For general situations where the axis of the filament is oriented with an viewing angle $\phi$ from the LOS, the rotation velocity along the LOS becomes `$v_{\rm rot}\sin\phi$' and the length differential becomes `$d/(\cos^2\theta\sin\phi){\rm d}\theta$'. Two `$\sin\phi$'s cancel each other and $\delta T_{\ksz}(d)$ remains the same as equation (\ref{eq:ksz_profile}).

\subsubsection{The density and rotation velocity profiles of filaments }
\begin{table}
 \caption{The average gas density in filaments in units of the cosmic-mean baryon density $\alpha_{\rm gas}$, measured from the hydrodynamical simulations of Z21. We refer readers to Figs.~5, 7 and Table 1 of Z21 for details.}
 \label{tab:alpha_gas}
 \centering
 \begin{tabular}{lcccc}
  \hline
  z & 0.0 & 0.5 & 1.0 & 2.0 \\
  \hline
  $\alpha_{\rm gas}$ & 6.5 & 6.0 & 5.7 & 4.4\\
  \hline
 \end{tabular}
\end{table}

{To estimate $\delta T_{\rm kSZ}(d)$, both $n_{\rm e}(r)$ and $v_{\rm rot}(r)$ are needed. For the density profile, we adopt the single-$\beta$ baryon density profile of filaments fitted from hydro-simulation measurements in Z21,
\begin{eqnarray}
\label{eq:den_prof}
	\rho_{\rm gas}(r,z)=
	\left\{
	\begin{array}{cc}
		\rho_{\rm gas,0}(z)\times\left[1+\left(\frac{r}{r_c}\right)^2\right]^{-\frac{3}{2}\beta_{\rm gas}}\,,& r\le R_{\rm fila}  \\
		\rho_{\rm gas,0}(z)\times\left[1+\left(\frac{R_{\rm fila}}{r_c}\right)^2\right]^{-\frac{3}{2}\beta_{\rm gas}}\,,& r> R_{\rm fila}
	\end{array}
	\right.
\end{eqnarray}
Here $r_c = 0.61R_{\rm fila}$, $R_{\rm fila}$ is the baryonic radius of filaments\footnote{It can be ambiguous how to define the radius of a filament. \cite{Zhu2021} follow the definition and measurement in \citealt{Cautun2014}, which refers to the local radius of a filament segment.}, $\beta_{\rm gas} = 2/3$, $\rho_{\rm gas,0}$ is the comoving baryon density at the filament spine, and $\rho_{\rm gas,0}=\alpha_{\rm gas}\bar{\rho}_{\rm baryon}$, where $\bar{\rho}_{\rm baryon}$ is the cosmic mean baryon density.
$\alpha_{\rm gas}$ depends on redshift, as illustrated by Figs.~5 and 7 of Z21. The fitted $\alpha_{\rm gas}$ values at four redshifts are listed in Table \ref{tab:alpha_gas}. We interpolate between these to calculate $\alpha_{\rm gas}(z)$ at the target redshift.} 

Furthermore, we assume that all baryons are fully ionised and the mean particle weight per electron $\mu_{\rm e} = 1.17$ \footnote{\url{https://www.ucolick.org/~woosley/ay112-14/texts/glatz.pdf}}. The number-density profile of free electrons is  $n_{\rm e}(r) = \rho_{\rm gas}(r)/(\mu_{\rm e}m_{\rm p})$, where $m_{\rm p}$ is the proton mass. The comoving $n_{\rm e}(r)$ at four different redshifts are plotted on the top panel of Fig.~\ref{fig:prof}. As shown, the density profiles are self-similar and the redshift-dependent amplitude is determined by $\alpha_{\rm gas}(z)$.  The profiles have a transition at $R_{\rm fila}$, beyond which we reach walls and voids, and the density profiles flatten out.

For the rotational velocity profile, we assume that the rotation of baryons follows that of dark matter. This may not be valid near the spine of the filament where the gas density is high and it may be pressure supported, but it should be a reasonable assumption further away from the spine. X21 has measured the dark matter rotational velocity profiles around filaments connecting pairs of dark matter halos (panel \textit{d} in Fig.~2 of X21). For their simulation setup, the rotation velocity of dark matter around filaments varies slowly within $r<2\mpch$, and then declines about linearly for $r\gtrsim 2\mpch$. Motivated by these simulation results, we model $v_{\rm rot}(r)$ with the following function,
\bea
\label{eq:vel_prof}
    v_{\rm rot}(r,z)= \alpha_{\rm curl}(z)\times
	\left\{
	\begin{array}{cc}
		85 {\rm km/s}\,,& r\le 2\mpch  \\
		\left(-25\frac{r}{\mpch}+135\right) {\rm km/s}\,. & r> 2\mpch
	\end{array}
	\right.
\eea
Here $\alpha_{\rm curl}(z)$ is introduced to account for the evolution of the rotational velocity. The power spectrum of the curl component of the velocity field evolves $\propto (afH)^2D^7$ (e.g.  \citealt{Pueblas09,Zheng13}, note that \citealt{Cristian2021} found a larger power index of $\propto D^{7.7}$). As part of the curl velocity field, we assume that the amplitude of the rotational velocity profile of filaments to evolve with the factor $\alpha_{\rm curl}(z)\propto (afH)D^{7/2}$ which is normalized to 1 at $z=0$. Fig.~2 of \citealt{Zhu2015} and Fig.~16 of \citealt{Zhu2017} have also measured the averaged curl velocity in different cosmic web environments at several redshifts. Our model of equation (\ref{eq:vel_prof}) is in qualitative agreement with those measurements from hydro simulations. We note that it could be a crude assumption that filament rotation velocities speed up with cosmic time according to the vorticity power spectrum amplitude. But we do expect the amount of rotation in filaments to grow roughly like this.

Inserting the above density and rotational velocity profiles into equation (\ref{eq:ksz_profile}),
$\delta T_{\ksz}(d)$ of a rotating filament can be calculated either analytically or numerically. We list the analytical expressions of equation (\ref{eq:ksz_profile}) in Appendix \ref{append:analytical_kSZ_profile}, and we have confirmed that these two ways yield consistent results. Following this, the angular distribution of $\delta T_{\ksz}(\theta_y)$ on a filament's projection plane can be obtained by replacing $d$ with $D_A\theta_y/a$ in equation (\ref{eq:ksz_profile}). $D_A$ is the angular diameter distance to the filament.

Finally, we apply a cut-off for the range of integration, $R_{\rm lim}$. We based our choice on the convergence tests shown in Appendix \ref{append:R_lim_convergence}. We have found that the S/N of the kSZ signal from the spin of filaments generally converges at $R_{\rm lim}=3R_{\rm fila}$, and it has reached $90\%$ of the total by $R_{\rm lim}=2R_{\rm fila}$. Therefore, we make a conservative cut at $R_{\rm lim}=2R_{\rm fila}$ by default. This should not affect the main conclusions of this paper.

\subsubsection{The filtered kSZ signal}
\begin{figure}
\begin{center}
\includegraphics[width=\columnwidth]{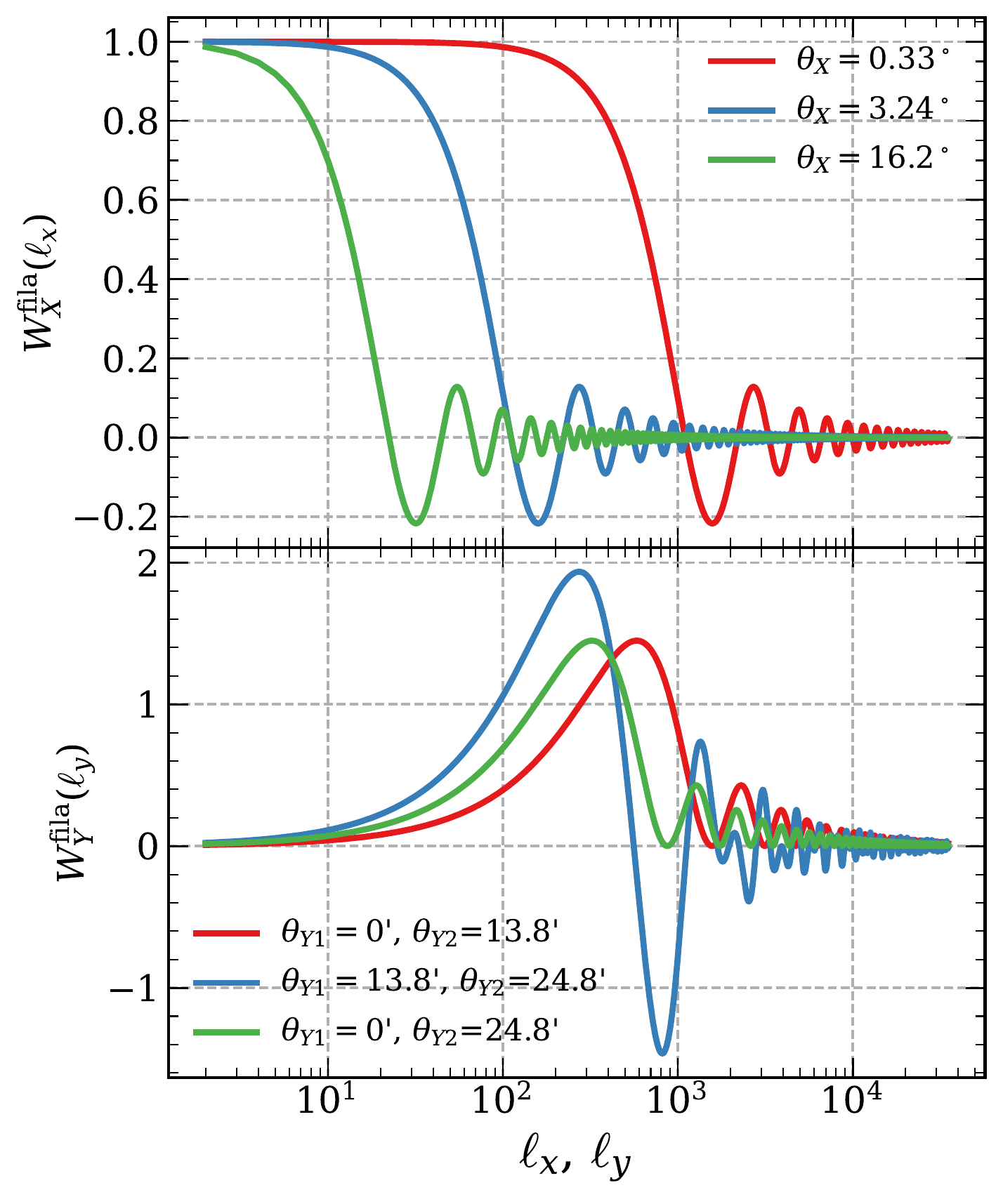}
\end{center}
\caption{Illustration of the filter function $W^{\rm fila}(\vec{\ell})$ applied to extract the kSZ signal from diffuse gas rotating around filaments. {$\theta_X$'s and $\theta_Y$'s represent the lengths and widths of filaments, chosen to be at $z=0.085$, which is the mean redshift of cold filaments in the Wang21 catalog.} $\theta_X = 0.33^\circ$, $3.24^\circ$ and $16.2^\circ$ correspond to filament lengths of $L_X = 1.43$, $14.1$ and $70.5$ Mpc/$h$. $\theta_Y = \theta_{Y2} - \theta_{Y1} = 13.8'$ and $24.8'$ correspond to $L_Y = 1$ and $1.8$ Mpc/$h$. They are shown in different colours.}
\label{fig:window}
\end{figure}
With the setup of the profiles, filter and noise from the previous subsections, we can now calculate the observed kSZ signal. As shown in Fig.~\ref{fig:filter}, the projection of the filament on the sky is considered to be a rectangle, and is equally divided into regions A and B by the filament spine. 

We measure the kSZ signal with $\delta T^{\rm fila}_{\rm kSZ} = \delta T^A_{\rm kSZ} -\delta T^B_{\rm kSZ}$. It is 
\beq
\label{eq:T_kSZ}
\delta T^{\rm fila}_{\rm kSZ}({\vec{\theta}}) = \int d^2\theta'W^{\rm fila}({\vec{\theta}-\vec{\theta}'})\delta T^{\rm obs,fila}_{\rm kSZ}(\vec{\theta}')\,,
\eeq
where $\delta T_{\rm kSZ}^{\rm obs,fila}$ is evaluated by equation (\ref{eq:T_ksz_obs}). By setting the center of the filament at $\vec{\theta}=0$, the normalized filter function is
\bea
W^{\rm fila}(\vec{\theta}) = \left\{
\begin{array}{cc}
 1/(\theta_X\theta_Y),\,\,\, |\theta_x| \leq \theta_{X}/2 \,\,{\rm and} \,\,\theta_{Y1} <\theta_y \leq \theta_{Y2} \\ \no
 -1/(\theta_X\theta_Y),\,\,\, |\theta_x| \leq \theta_{X}/2 \,\,{\rm and}\,\, -\theta_{Y2} \leq \theta_y\leq -\theta_{Y1} \\ \no
 0,\,\,\, {\rm elsewhere}\,. \no
\end{array}\right.
\eea
Here $\theta_X$ corresponds to the length of the filament, and $\theta_Y = \theta_{Y2} - \theta_{Y1} \le aR_{\rm fila}/D_A$ is the width of the region within which the measurement is taken. Its Fourier transform is:
\bea
W^{\rm fila}(\vec{\ell}) &=& \int d^2\theta e^{-i\vec{\ell}\cdot\vec{\theta}}W^{\rm fila}(\vec{\theta}) \no \\
&=& 2i\sin[(\theta_{Y1}+\theta_{Y2}) \ell_y/2] {\rm sinc}(\theta_X\ell_x/2){\rm sinc}(\theta_Y\ell_y/2) \no \\
&\equiv& i W^{\rm fila}_X(\ell_x)W^{\rm fila}_Y(\ell_y)\,,
\label{eq:w_l}
\eea
where 
\bea
W^{\rm fila}_X(\ell_x)&\equiv&{\rm sinc}(\theta_X\ell_x/2)\,,\\
W^{\rm fila}_Y(\ell_y)&\equiv& 2\sin[(\theta_{Y1}+\theta_{Y2}) \ell_y/2]{\rm sinc}(\theta_Y \ell_y/2)\,.
\eea

Fig.~\ref{fig:window} shows examples for the shape of $W^{\rm fila}(\vec{\ell})$. The chosen $\theta_X$'s and $\theta_Y$'s represent filaments at $z=0.085$, which is the mean redshift of the filament catalog found in W21 (hereafter Wang21 catalog). As shown in the figure, $W^{\rm fila}_X$ damps the small scale (high-$\ell$) CMB fluctuations and keeps the cosmic variance on large scales; $W^{\rm fila}_Y$ suppresses the large-scale (low-$\ell$) {and small- scale (high-$\ell$)} perturbation modes. Their combination simultaneously damps large and small scale fluctuations, keeping the kSZ signal mainly at the intermediate scales.

\subsection{kSZ signal from galaxies rotating around filaments}

An alternative way of detecting the spin is to measure the kSZ in small apertures around each galaxy in a filament. In this case, the kSZ signal of a galaxy can be obtained through a circular AP filter $W^{\rm gal}(\theta)$ (e.g.  \citealt{Sugiyama18,Calafut2021,Chen2022}),
\begin{eqnarray}
	W^{\rm gal}(\theta)=\frac{1}{\pi\theta^2_{\rm c}}
	\left\{
	\begin{array}{cc}
		1,&\quad \theta\leq\theta_{\rm c}  \\
		-1,&\quad \theta_{\rm c} < \theta \leq \sqrt{2}\theta_{\rm c}\\
		0,&\quad \theta > \sqrt{2} \theta_{\rm c} 
	\end{array}
	\right.
\end{eqnarray}
in which the averaged CMB temperature of pixels within a disk of aperture size $\theta_{\rm c}$ and an annulus of equal area, out to radius $\sqrt{2} \theta_{\rm c}$, are differenced around each galaxy. Therefore, the kSZ signal of a galaxy is
\beq
\delta T^{\rm fila,gal}_{\rm kSZ}({\vec{\theta}}) = \int d^2\theta'W^{\rm gal}({\vec{\theta}-\vec{\theta}'})\delta T^{\rm obs,gal}_{\rm kSZ}(\vec{\theta}')\,,
\eeq
where $\delta T_{\rm kSZ}^{\rm obs,gal}$ is evaluated by equation (\ref{eq:T_ksz_obs}). 

We assume that the CMB photons are scattered off by the free electrons of a single galaxy before they reach the observer. The kSZ signal of a galaxy is thus reduced to 
\beq
\label{eq:kSZ_oneobj}
\delta T_{\rm kSZ}(\hat{n}_i) = - \frac{T_0\tau_{{\rm T},i}}{c}\bfv_i\cdot\hat{n}_i \,,
\eeq
in which $\tau_{{\rm T},i}=\int dl\sigma_{\rm T} n_{e,i}$ is the optical depth of the electron cloud associated with the $i$th galaxy.

By further assuming that all gas halos of galaxies have the same mass, density profiles, and thus the same total $\tau^{\rm gal}_{\rm T}$, we can now difference the averaged kSZ signal for each galaxy on the A and B regions. This yields the estimated kSZ signal associated with galaxies rotating around filaments:
\beq
{\delta T^{\rm fila,gal}_{\rm kSZ} = \frac{\sum_l \delta T^{A,\rm gal}_{{\rm kSZ},l}}{\sum_l}-\frac{\sum_m \delta T^{B,\rm gal}_{{\rm kSZ},m}}{\sum_m} = -2\frac{T_0\tau^{\rm gal}_{\rm T}}{c}\frac{\int\bar{v}_{||}(r){\rm d}r}{\int {\rm d}r}\,,}
\label{eq:ksz_gal}
\eeq
where $l,m$ denote the $l$th and $m$th galaxy in region-A and -B separately. $\bar{v}_{||}(r)$ is the averaged rotation velocity of a galaxy projected along the LOS of the region-A\footnote{We actually observe galaxies in redshift space. The filament rotation will transfer the filament galaxies between different $r$ bins in redshift space. However, as long as we observe all galaxies associated with a filament in redshift space, this effect will not bias the signal described by equations (\ref{eq:ksz_gal}) and (\ref{eq:v_par}).}, which can be calculated by
\beq
\label{eq:v_par}
\bar{v}_{||}(r) = \frac{\int_0^{\pi/2} v_{\rm rot}(r)\cos\theta {\rm d}\theta}{\int_0^{\pi/2}{\rm d}\theta}=\frac{2}{\pi}v_{\rm rot}(r) \,.
\eeq

Furthermore, if we have a filament sample with an orientation distribution $f_i = f(\phi_i)$, the orientation will affect the $v_{\rm rot}$ along the LOS and the average over the filament sample is
\beq
\label{eq:v_par_prime}
\langle\bar{v}_{||}(r)\rangle = \bar{v}_{||}(r)\sum_i \sin(\phi_i)f_i \equiv \alpha_\phi \bar{v}_{||}(r)\,.
\eeq
In turn, the orientation-averaged kSZ signal from galaxies rotating around filaments is
\beq
\delta T^{\rm fila,gal}_{\rm kSZ} = -\frac{4T_0\tau^{\rm gal}_{\rm T}\alpha_\phi}{\pi c}\frac{\int v_{\rm rot}(r){\rm d}r}{\int {\rm d}r}\,.
\label{eq:ksz_gal_avg}
\eeq
We estimate $\alpha_{\phi}= 0.81$ from the `dynamically cold' filament orientation distribution shown in the bottom panel of Fig. \ref{fig:hist}. Here the `coldness' of a filament is defined by $z_{\rm rms}/\Delta z_{\rm AB}$, in which $z_{\rm rms}$ is the root mean square of the galaxy redshifts inside this filament, and the dynamically cold filaments are those with $z_{\rm rms}/\Delta z_{\rm AB} < 1$. We refer readers to Appendix~\ref{append:wang21_catalog} for details of this filament sample.

For all the calculations in this work, the filament galaxies are set to reside in regions with $r<2\mpch$ (see Fig.~ \ref{fig:hist} for details), where the rotation velocity is assumed to be constant ($v^{\rm const}_{\rm rot}$), therefore equation (\ref{eq:ksz_gal_avg}) reduces to 
\beq
\delta T^{\rm fila,gal}_{\rm kSZ} = -\frac{4T_0\tau^{\rm gal}_{\rm T}\alpha_\phi}{\pi c}v^{\rm const}_{\rm rot}\,.
\label{eq:ksz_gal_avg_vcostant}
\eeq

Finally, the Fourier transform of the circular AP filter $W^{\rm gal}(\theta)$ is \citep{Alonso2016}
\beq
\label{eq:cir_ap}
W^{\rm gal}(x) = 4\frac{J_1(x)}{x}-4\frac{J_1(\sqrt{2}x)}{\sqrt{2}x}\,,
\eeq
where $x = \ell \theta_{\rm c}$ and $J_1(x)$ is the first Bessel function of the first kind. As shown in Fig. \ref{fig:cir_window}, this is a compensated window function in 2-D. By inserting equation (\ref{eq:cir_ap}) into equation (\ref{eq:noise}), we can calculate the kSZ statistical noise associated with a galaxy. The last ingredient needed for the above predictions is a model for the optical depth of the gas halo around galaxies, which we will address in the next sub-section.

\subsubsection{The optical depth}
\begin{figure}
\begin{center}
\includegraphics[width=\columnwidth]{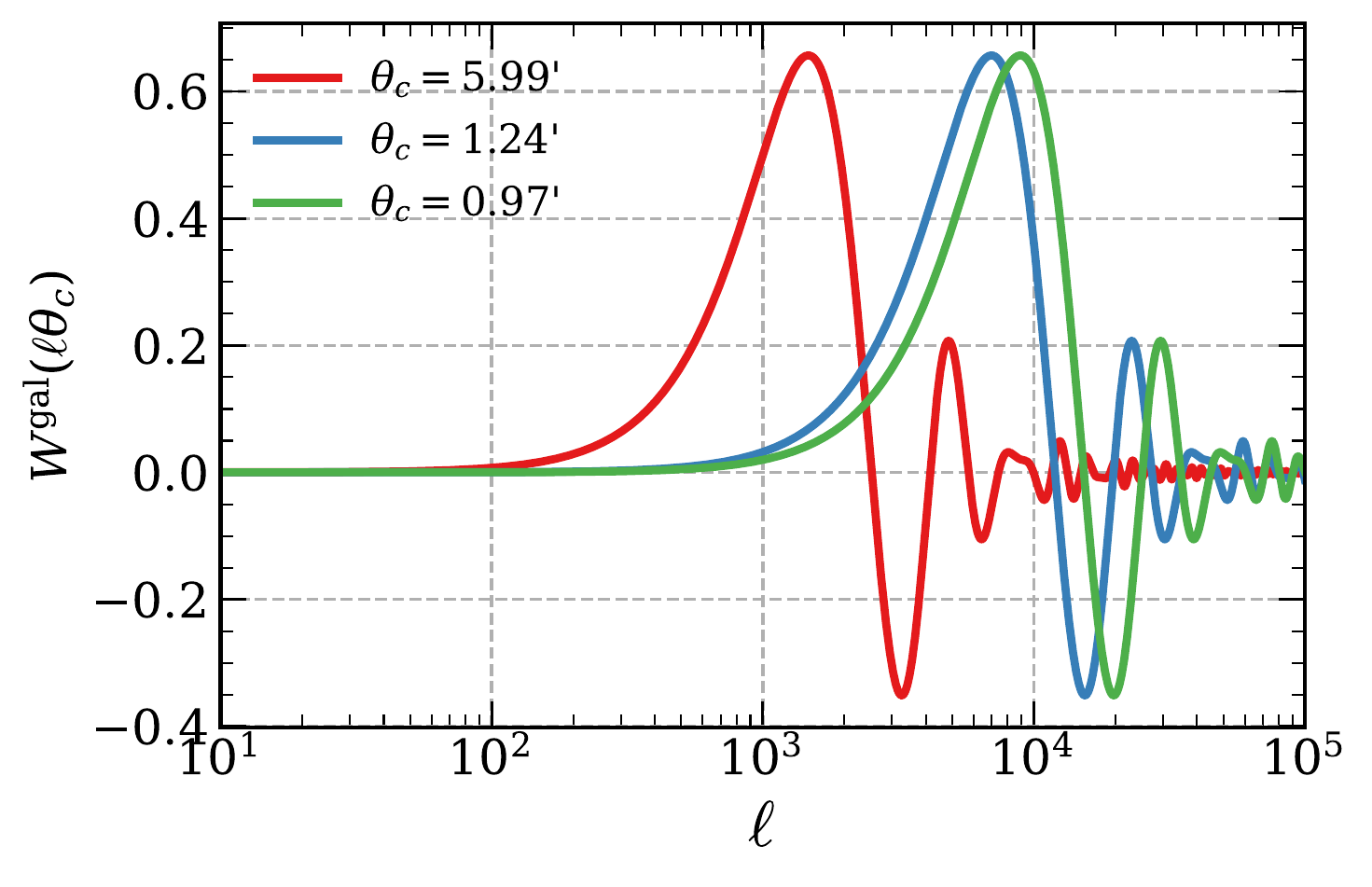}
\end{center}
\caption{Illustration of the circular AP filter $W^{\rm gal}(\ell\theta_{\rm c})$. $\theta_{\rm c}$ is the optimised aperture radius in equation (\ref{eq:theta_c_peak}) at $z=0.1$, $0.9$ and $1.9$ of a $M_h=5\times10^{12}$M$_\odot/h$ which maximizes the optical depth.}
\label{fig:cir_window}
\end{figure}
\begin{figure}
\begin{center}
\includegraphics[width=\columnwidth]{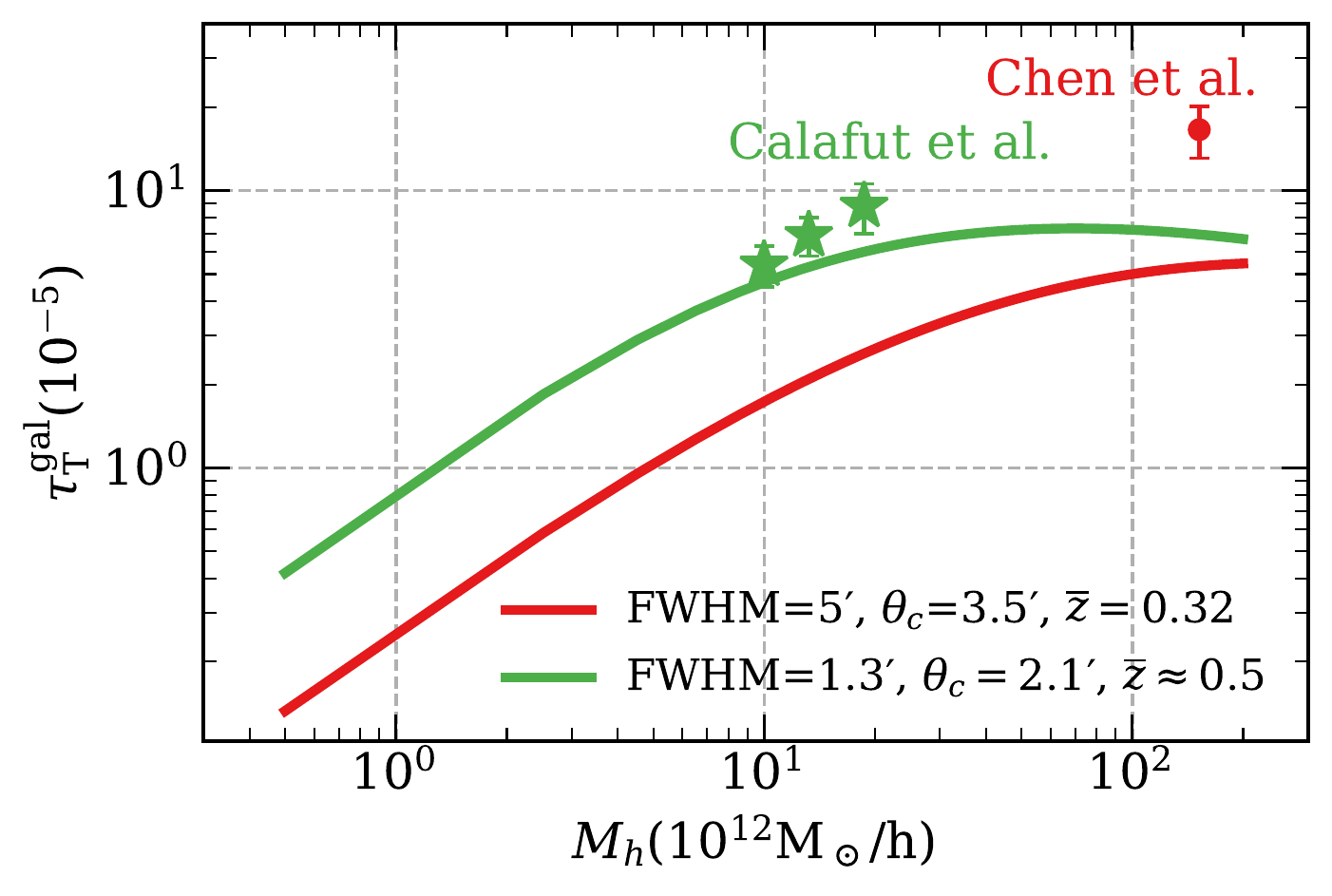}
\end{center}
\caption{Comparison of the modeled optical depth $\tauT$ versus halo mass from equation (\ref{eq:tau_T}) and measurements from observations of \citealt{Chen2022} (red data points with errors) and \citealt{Calafut2021} (green data points with errors). The two different colour lines shows the theoretical calculations with different assumptions for the FWHM for the CMB, filter aperture, and mean redshift of the halo, as labeled in the legend.}
\label{fig:tauT_M}
\end{figure}

For a halo containing gas with a projected profile $N(\vec{\theta})$, the optical depth can be calculated with \citep{Sugiyama18}
\beq
\tauT = \frac{\sigma_{\rm T}f_{\rm gas}M_{\rm h}}{\mu_{\rm e}m_{\rm p}D^2_{\rm A}(z)}\int \frac{{\rm d}^2\ell}{{2\pi}^2}W^{\rm gal}(\ell\theta_{\rm c})N(\vec{\ell})B(\vec{\ell})\,,
\label{eq:tau_T}
\eeq
where $N(\vec{\ell})$ is the Fourier transform of $N(\vec{\theta})$. The value of $\tauT$ varies with many factors: the assumed $N(\vec{\theta})$, $f_{\rm gas}$, halo mass $M_{\rm h}$, CMB beam size $\theta_{\rm B}$ and the adopted $\theta_{\rm c}$. For simplicity, we adopt an universal gas-mass fraction $f_{\rm gas}=\Omega_b/\Omega_m=0.158$~\citep{Lim2020} and a Gaussian projected gas profile,
\beq
N(\vec{\theta}) \propto \frac{1}{2\pi\sigma_{\rm R}^2} {\rm e}^{-\theta^2/(2\sigma_{\rm R}^2)}\,
\eeq
in the calculation, where the typical gas distribution scale $\sigma_{\rm R}=R_{\rm 200c}/D_{\rm A}$. We compare the $\tauT$'s calculated from equation (\ref{eq:tau_T}) with measurements from observations from \citet{Calafut2021} and \citet{Chen2022} in Fig. \ref{fig:tauT_M}.

The measurements of \citealt{Calafut2021} come from the analyses of SDSS DR15 data and the ACT data; while \citealt{Chen2022} combines the DESI imaging survey with the Planck data \citep{Chen2022}. These two measurements represent the gas halos associated with galaxies and galaxy clusters respectively. As shown in the figure, despite the simplifications we have made in the model, the predicted optical depths agree reasonably well with those of \citealt{Calafut2021}, but slightly under predict that of \citealt{Chen2022}. We will mainly consider halos with mass $M_h\leq 10^{13}{\rm M}_{\odot}/h$ in this work, thus the slight under-prediction of the model at $M_h\sim 10^{14}{\rm M}_{\odot}/h$ is not a major problem, and if the model is indeed under-predicting $\tauT$ at the high-mass end, it will lead to a lower S/N for the kSZ signal. So one can take our calculations as a lower-limit. 

With the assumptions that both the beam profile and the gas profiles of halos are Gaussian, we can adopt an optimised aperture radius which maximises the optical depth \citep{Sugiyama18}
\beq
\label{eq:theta_c_peak}
\theta_{\rm c} = 2^{3/4}\sqrt{\sigma_{\rm B}^2+\sigma^2_{\rm R}}\,,
\eeq
which is similar to applying a matched-filtering technique in the kSZ signal extraction (e.g.  \citealt{Alonso2016}). The associated optical depth of the halo can be approximated by 
\bea
\label{eq:tau_T_halo}
\tauT^{\rm gal} &=& 5.37\times 10^{-5} \left(\frac{f_{\rm gas}}{0.158}\right)\left(\frac{M_h}{10^{14}h^{-1}M_\odot}\right) \no\\
&&\times \left(\frac{h}{0.68}\right)\left(\frac{3'}{\sqrt{\sigma_{\rm B}^2+\sigma_{\rm R}^2}}\right)^2\left(\frac{10^3h^{-1}{\rm Mpc}}{D_{\rm A}}\right)^2\,.
\eea
Here $\sigma_{\rm R} = R_\Delta/D_{\rm A}$ denotes a characteristic radius of the projected Gaussian gas profile $N(\vec{\theta})$, and $R_\Delta$ is the physical halo radius within which the average density of the halo is $\Delta$ times of the critical density $\rho_{\rm crit}(z)$ at redshift $z$,
\beq
R_\Delta = \left(\frac{3}{4\pi}\frac{M_h}{\Delta \rho_{\rm crit}(z)}\right)^{1/3} \,,
\eeq
with $\Delta = 200$.

\section{Forecast for current and future surveys}
\label{sec:ston_estimation}

In this section we estimate the S/N for the kSZ signal from the spin of filaments from the current and future observations. We first apply the theoretical framework to the Wang21 catalog from SDSS, where a tentative detection for the spin of filaments was found (W21). We then focus our forecast on the on-going galaxy survey DESI \citep{DESI2016} and future 21 cm surveys, e.g. SKA-2 \citep{Maartens2015} and their combination with future CMB surveys.

 Throughout the calculations, we adopt the flat $\Lambda$CDM model with the PLANCK2018 \citep{PLANCK2018} best-fit parameters as the fiducial cosmological model: $\Omega_{\rm m}=0.3111$, $\Omega_{\rm b}=0.049$, $n_{\rm s}=0.9665$, $A_{\rm s}=2.105\times10^{-9}$ and $h=0.6766$. To estimate the detection noise from equation (\ref{eq:noise}), we generate the lensed CMB temperature power spectrum by running the \textsc{camb} code \citep{CAMB}. The calculations of cosmological quantities are done via the \textsc{colibri}\footnote{\hyperlink{ https://github.com/GabrieleParimbelli/COLIBRI}{ https://github.com/GabrieleParimbelli/COLIBRI}} python package.

\subsection{Wang21 catalog}
\label{subsec:ston_estimation_W21}
First, we estimate the S/N of the kSZ signal imprinted on the Planck temperature maps for a sample of filaments from W21. We choose the 217 GHz frequency map, which has the resolution of ${\rm FWHM} = 5'$ and the detector noise of $\sim46.8 \muk$-$\rm arcmin$ \citep{Planck2016}. The detailed description of Wang21 catalog can be found in appendix \ref{append:wang21_catalog}. A sub-sample of this catalog was selected in W21 such that the spin induced redshift difference between the region-A and -B is larger than the redshift dispersion of galaxies within a filament. This criteria makes sure that the blue- and redshift of galaxies in the two regions are well separated from the noise, i.e. having a relatively clean redshift dipole for the galaxies. We denote this sub-sample of filaments as a `dynamically cold' sample hereafter. This cold sample has $N^{\rm cold}_{\rm fila}=5964$ filaments, with $N^{\rm cold}_{\rm fila,gal}=54417$ member galaxies. We will focus on the S/N prediction for this cold sample in the following.

\subsubsection{kSZ signal from diffuse gas rotating around filaments}
\label{subsubsec:deltaT_fila}
\begin{figure}
\centering
\includegraphics[width=\columnwidth]{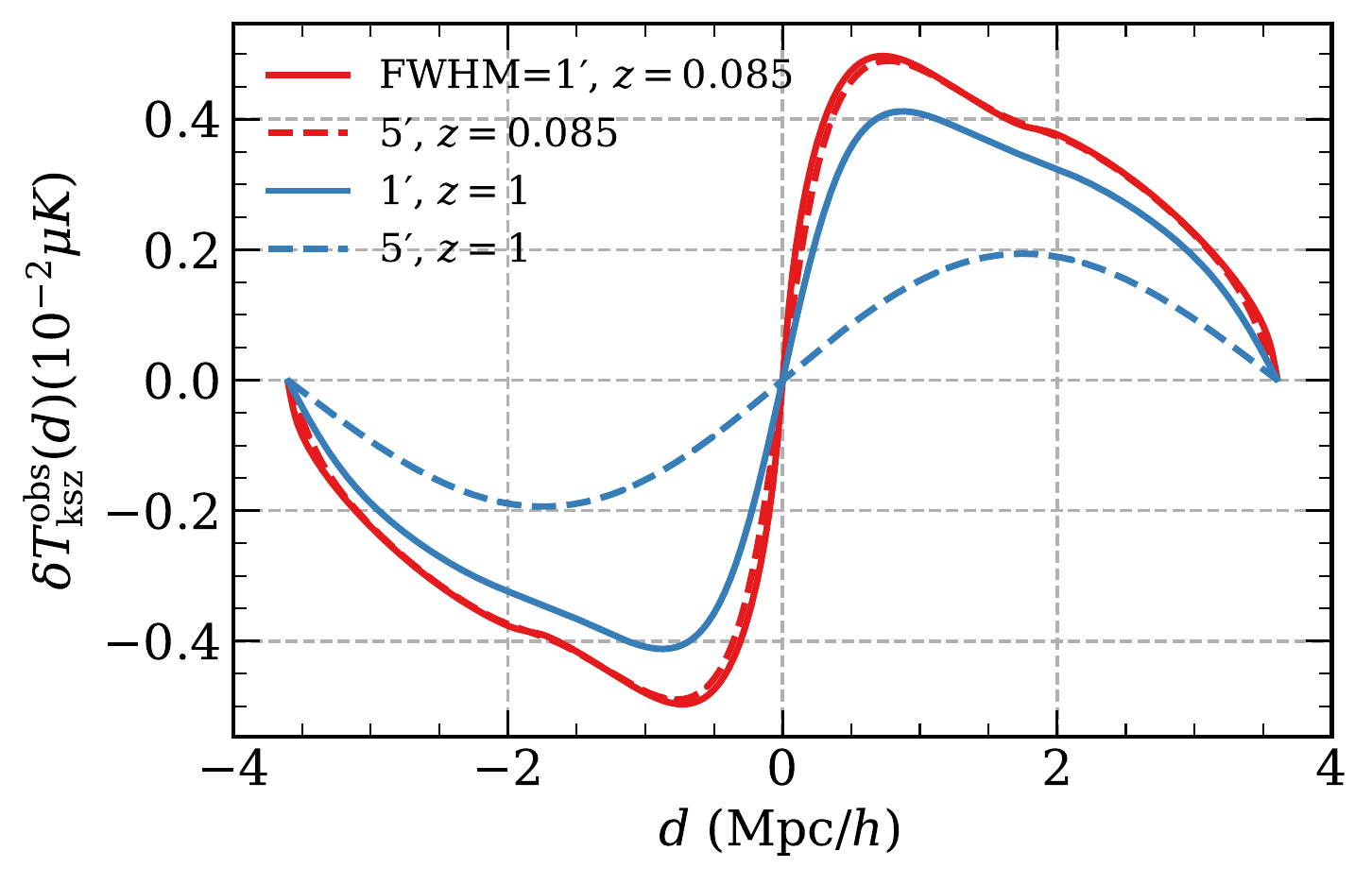}
\caption{The kSZ temperature profiles expected from the cold sample of Wang21 catalog calculated by equation (\ref{eq:ksz_profile}) and the associated expressions in Appendix \ref{append:analytical_kSZ_profile}. $d$ is the distance perpendicular to the spine of the filament. The radius of the filament is taken to be $R_{\rm fila} =1.8 \mpch$ and we integrate up to $R_{\rm lim}=2R_{\rm fila}$. The FWHM adopted for the beam of the CMB and the redshift for the filament are labeled in the legend.}
\label{fig:TkSZ_dipole_wang21}
\end{figure}
\begin{figure}
    \centering
    \includegraphics[width=\columnwidth]{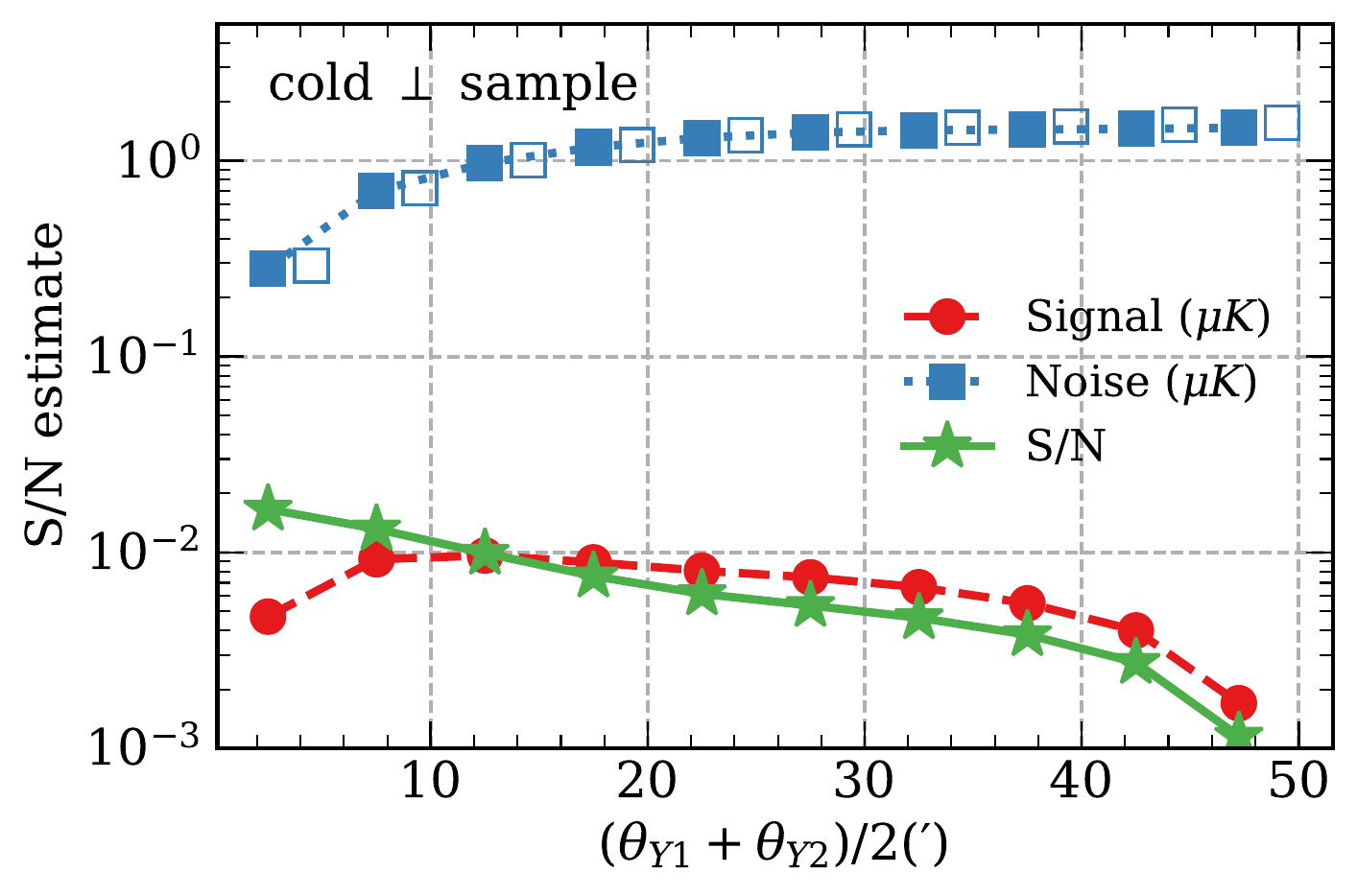}
    \caption{The expected filtered temperature of the kSZ signal, noise and S/N from the cold $\perp$ sample of Wang21 catalog and Planck data. They are plotted against the width of the filter $(\theta_{Y1}+\theta_{Y2})/2$. $\theta_{Y1} = (0,5',...,40',45')$, $\theta_{Y2} = (5',10',...,45',49.5')$ and the length of the filament is set to be $\theta_X=3.24^\circ$. The blue empty squares next to the solid ones are shifted to the right by $1'$ for better illustration. They represent the noise calculated from the observed Planck 217 GHz temperature map. The solid blue squares are for the same quantities calculated analytically from our noise model. Their good agreement indicate the robustness of our noise model. The temperature of kSZ signal from the rotation of diffuse filament gas is shown in red circles, and the S/N is shown in green stars, as indicated by the legend. There are 2048 filaments in this sub-sample, whose averaged redshift is $z_{\rm avg}^{\rm cold,\perp}=0.085$. Its details are presented in Appendix \ref{append:wang21_catalog}.}
    \label{fig:ston_fila_wang21}
\end{figure}
Filaments in the Wang21 catalog are found through the Bisous model \citep{Tempel2014}, a marked object point process with interactions, from the Legacy survey of SDSS DR12 data \citep{SDSSdr12}. Filament radius, denoted as $R_{\rm gal}$ in the Wang21 catalog, is defined by the galaxy number density. As galaxies tend to have relatively higher linear biases than the total baryonic matter, they are more clustered. Hence $R_{\rm gal}$ is usually smaller than the radius defined by the baryonic or DM density as in simulation studies (e.g.  Z21). Since the kSZ signal is closely related to the baryonic  distribution in the filament, we need to estimate the baryonic radius $R_{\rm fila}$ of the Wang21 catalog filaments. To do this, we use the relationship between the total length of filaments and their diameter, measured from hydro-dynamical simulations of Z21 (the top left panel of their Fig.~2 ). We first compute the total length of filaments per $({\rm Gpc}/h)^3$ in the Wang21 catalog, which is $1.78\times10^6 \mpch/({\rm Gpc}/h)^3$. Then we apply the above relationship to the Wang21 catalog and estimate its average filament baryon radius to be $R_{\rm fila} \approx 1.8\mpch$. As a result, $r_c = 0.8R_{\rm fila} \approx 1.44\mpch$, and $\alpha_{\rm gas} \approx 6.5$, inferred from Table \ref{tab:alpha_gas} and Fig.~5 of Z21. 

By substituting equations (\ref{eq:den_prof}) and (\ref{eq:vel_prof}) into equation (\ref{eq:ksz_profile}), and applying a Gaussian beam function in equation (\ref{eq:T_ksz_AP}), we estimate the kSZ signal from diffuse gas rotating around filaments in the Wang21 catalog, $\delta T^{\rm obs}_{\rm kSZ}(d)$, shown in Fig.~\ref{fig:TkSZ_dipole_wang21}. We can see that the signal profile is an odd function due to the anti-symmetric nature of the velocity profile. The amplitude of the kSZ signal decreases towards the centre due to the decreasing amplitude of the LOS component of the rotational velocity; it also decreases with increasing distance $d$ owing to the fact that the overall amplitude of the rotational velocity decreases with $d$.

The solid lines are the expected kSZ signal profile convolved with the ACT/SPT/CMB-S4-like CMB beam function (${\rm FWHM}=1'$), while the dashed lines show the results after convolving with a Planck-like beam function (${\rm FWHM} = 5'$). Since $R_{\rm lim}=2R_{\rm fila}=3.6\mpch$, it spans around $50'$ on the sky at $z=0.085$, the signal dilution due to the Planck-like beam has little difference from that of ${\rm FWHM} = 1'$. However, at $z=1$, the angular scale of the filament becomes much smaller ($\sim 5.4'$). The observed kSZ signal is strongly suppressed by the Planck-like beam, but it remains largely unchanged with a finer beam (${\rm FWHM}=1'$).
This highlights the importance of high-resolution CMB data for observing the rotation of high-$z$ filaments. 

There is little redshift evolution for the kSZ signal between these two redshifts. This is expected from Eq.~\ref{eq:ksz_profile}: the $a^{-2}$ factor cancels some of the evolution for the rotational velocity and density profiles shown in Fig.~\ref{fig:prof}. The amplitude of the kSZ signal associated with the spin of filament is at the order of $10^{-3} \muk$, a factor of over 1000 smaller than the expected kSZ signal associated with the rotation of galaxy clusters \citep{Chluba2002,Cooray2002, Baldi2018, Baxter2019}. This is because the gas density in a filament is typically one or two orders of magnitude lower than in galaxy clusters, and the rotational velocity about one order of magnitude smaller. Compared to a galaxy cluster, a filament typically has a larger aperture to average over, working in favor of rotation detection. Still, the small amplitude of the signal poses a challenge for its detection in real observations.

Next, we apply the rectangular aperture filter $W(\vec{\theta})$ for $\delta T^{\rm obs}_{\rm kSZ}(d)$ to yield $\delta T^{\rm fila}_{\rm kSZ}$. We separate $d$ into 10 bins, {with the beam centres are at $(\theta_{Y1}+\theta_{Y2})/2 = (2.5',7.5',...,42.5',47.2')$ }. The bin size $\Delta d$ of $\Delta \theta_Y = 5'$ is chosen to {reduce the correlations between different bins due to the convolution with the beam. The resultant $\delta T^{\rm fila}_{\rm kSZ}(\theta_{Y1,Y2})$ are shown in red solid circles and dashed line in Fig.~\ref{fig:ston_fila_wang21}. We can see again that the amplitude of the filtered kSZ signal is at the level of $10^{-3} \muk$. The decrease of the signal towards small and large size of the filter [($\theta_{Y1} + \theta_{Y2}$)/2] is consistent with the shape of the kSZ profile shown in Fig.~\ref{fig:TkSZ_dipole_wang21}.

In Fig.~\ref{fig:ston_fila_wang21}, we also present the S/N estimate for a cold $\perp$ sample with the viewing angle $\phi>70^\circ$. A similar sub-sample of filaments was selected in W21 to maximize the S/N of the redshift dipole.

There are $N^{\rm cold,\perp}_{\rm fila}=2048$ filaments in this sub-sample, with an average viewing angle $\approx90^\circ$. The statistical noise of each $\theta_{Y1,Y2}$ bin can be estimated as 
\beq
\label{eq:err_fila_per}
\sigma^{\rm fila,\perp}_{\rm N,mean}(\theta_{Y1,Y2}) = \sqrt{[\sigma_{\rm N}^{\rm fila}(\theta_{Y1,Y2})]^2/N^{\rm cold,\perp}_{\rm fila}}\,,
\eeq 
where $[\sigma_{\rm N}^{\rm fila}(\theta_{Y1,Y2})]^2$ is computed by combining equations (\ref{eq:noise}) and (\ref{eq:w_l}). We adopt $L_{\rm avg} = 14.1\mpch$ and $\phi=90^\circ$ to calculate $\theta_X=3.24^\circ$ here. The results are shown with blue solid squares and the dotted line in Fig. \ref{fig:ston_fila_wang21}. 

To validate the above calculations for the noise, we also estimate the noise directly measured from the observed Planck 217 GHz map. These are shown in blue empty squares next to the solid ones. They are manually shifted to the right by $1'$. The analytical results for $\sigma_{\rm N,mean}^{\rm fila,\perp}$ agree very well with the directly measured ones.

As we can see, the statistical noise is generally one or two orders of magnitude larger than the signal. Therefore, the S/N's, shown with green stars and solid line in Fig. \ref{fig:ston_fila_wang21} for this sample are low. When we combine 10 bins and ignore the correlations between them, the resultant S/N of the cold $\perp$ sample is
\bea
\label{eq:ston_fila_cold_per}
\left(\frac{S}{N}\right)_{\rm cold,\perp}=\sqrt{\sum_{i=1}^5 [\delta T^{\rm fila}_{\rm kSZ}(\theta^i_{Y1,Y2})/\sigma^{\rm fila,\perp}_{\rm N,mean}(\theta^i_{Y1,Y2})}]^2\\
\approx 0.028\,. \no
\eea
We also estimate the S/N for all $N_{\rm cold} = 5964$ cold filaments. 

We separate the sub-sample into 9 orientation bins with bin width $\Delta \phi=10^\circ$. For each $\phi$ bin, $\theta_Y$ is binned as before. The resultant S/N of the cold sample is
\bea
\left(\frac{S}{N}\right)_{\rm cold}=\sqrt{\sum_{j=1}^{9}\sum_{i=1}^{10} [\delta T^{\rm fila}_{\rm kSZ}(\theta^i_{Y1,Y2})/\sigma^{\rm fila}_{\rm N,mean}(\theta^i_{Y1,Y2},\theta^j_X)}]^2 \\
\approx 0.043\,. \no
\eea
Therefore, including cold filaments with all orientations will enhance the S/N roughly by a factor of $0.043/0.028 \approx 1.55$, {but the total S/N is still very low when having the catalog combined with Planck data. The size of the sample needs to be boosted by a factor of $\sim 10000$ to beat down the statistical noise and achieve a $3-4\sigma$ detection. Such a factor can of course be reduced when the beam and the detector noise of the CMB survey is reduced compared to Planck, which is generally the case for the next generation of CMB surveys. }

\subsubsection{{kSZ signal from galaxies spinning around filaments}}
\begin{figure}
\includegraphics[width=\columnwidth]{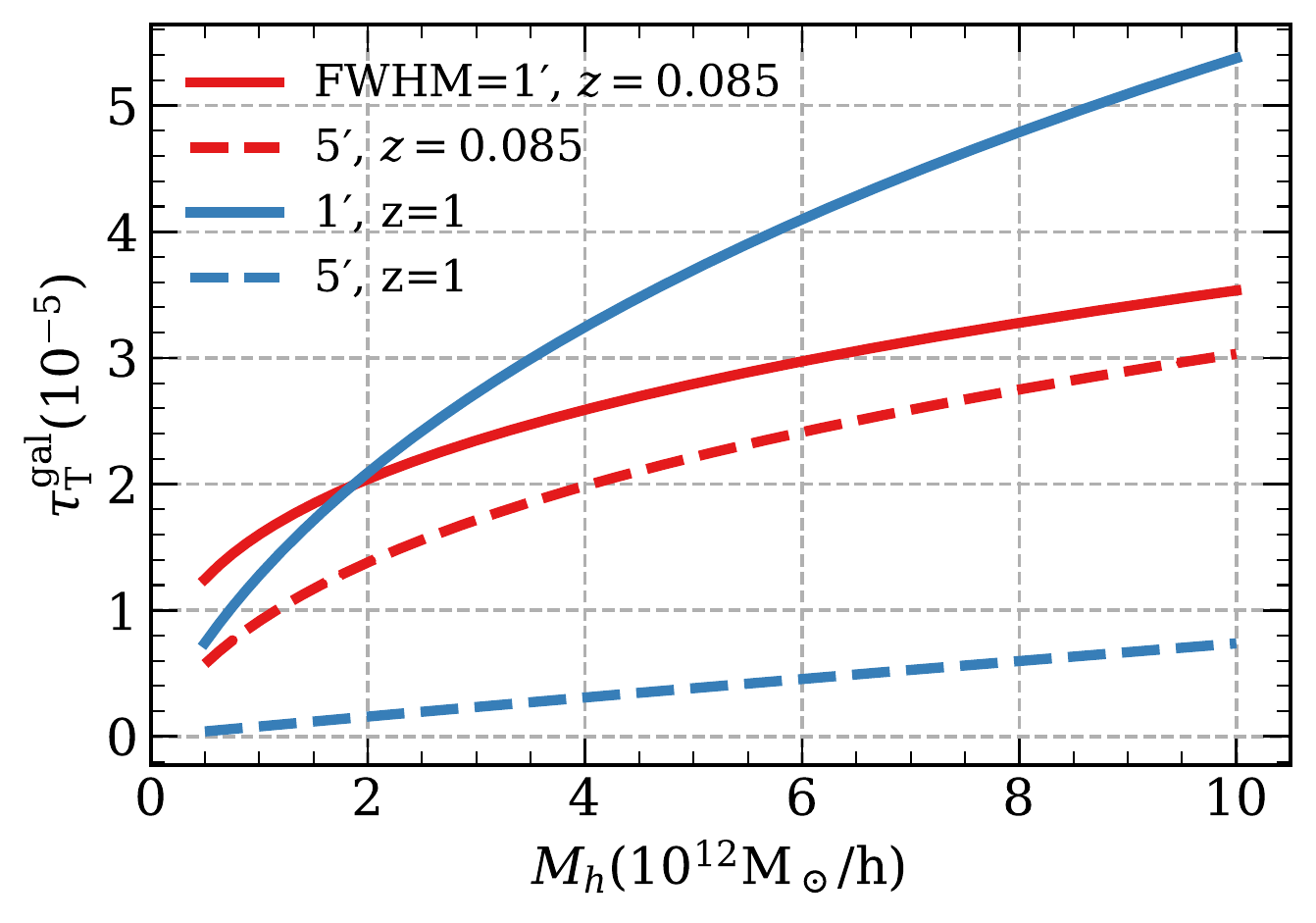}
\caption{{The estimated optical depth of a galaxy in a halo with the mass $M_h$ at two different redshifts and with two different FWHMs for the CMB as indicated by the legend.}}
\label{fig:tauT_wang21}
\end{figure}
\begin{figure}
    \centering
    \includegraphics[width=\columnwidth]{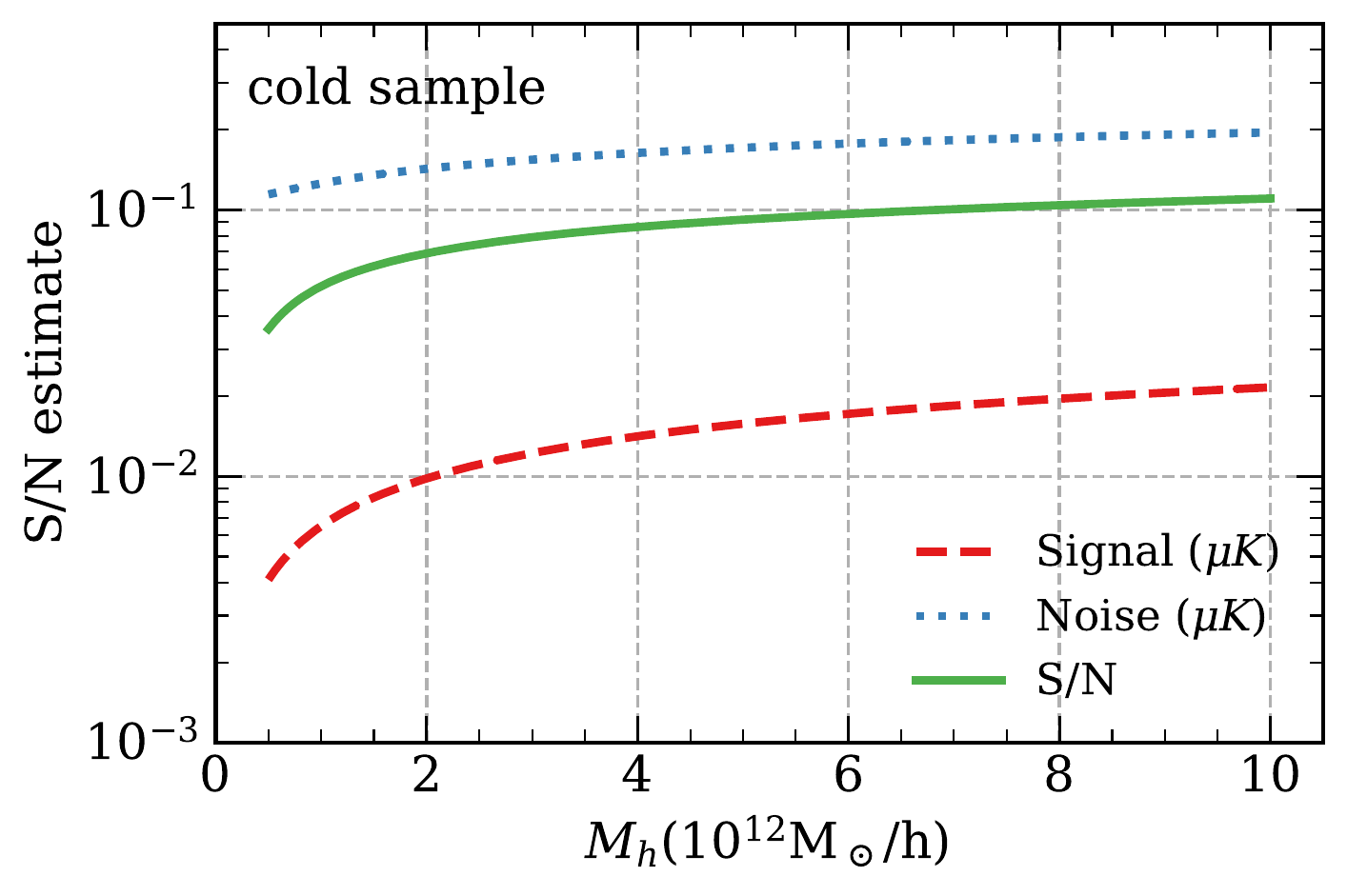}
    \caption{The S/N of the kSZ signal associated with the rotation of galaxies around filaments expected from the cold sample of filaments from W21 and Planck data. It is plotted against the assumed host halo mass $M_h$. There are 54417 filament galaxies in this sub-sample, whose averaged redshift is $z_{\rm avg}^{\rm cold} = 0.085$. Its details can be found in Appendix \ref{append:wang21_catalog}.}
    \label{fig:ston_fila_gal_wang21}
\end{figure}

Next we evaluate the S/N of the {kSZ signal associated with galaxies spinning around filaments}. {The host halo mass $M_h$ of galaxies around filaments is the main variable for this calculation. We consider halos of $0.5-10\times10^{12}M_{\odot}/h$ \citep{Cautun2014}.}

Fig. \ref{fig:tauT_wang21} shows the optimised optical depth of a halo as a function of the halo mass. As expected, halos with larger mass contain more free electrons and have larger $\tauT$. Since the celestial extend of a galaxy is relatively small ($R_{200c}/D_A \sim 5'$ for a $M_h=10^{12}M_\odot/h$ halo at $z=0.085$), the large Planck-like ${\rm FWHM}=5'$ notably degrades the $\tauT$, as shown by comparing the solid and dashed red lines in Fig. \ref{fig:tauT_wang21}. This degradation is much more severe at higher redshifts (e.g.  $z=1$ in Fig. \ref{fig:tauT_wang21}) due to the smaller halo celestial extend. Hence we notice again that the finer CMB instrumental resolution is necessary for the kSZ signal extraction at high redshifts.

With equation (\ref{eq:ksz_gal_avg}), we estimate the expected kSZ signal from galaxies spinning around filaments for all the 54417 galaxies in the cold sample. The results are shown by the red dashed line in Fig.~\ref{fig:ston_fila_gal_wang21}. Consistent with the previous plot, the signal increases as the halo mass becomes larger. We also notice that the signal does not increase linearly with $M_h$,  as seemingly shown in equation~(\ref{eq:tau_T_halo}). This is because the $\sigma_{\rm R}$ therein also increases with $M_h$ and slows down the overall enhancement of $\tauT$ with a larger $M_h$.

The estimated errors of the mean for the kSZ signal are estimated with
\beq
\label{eq:err_filagal}
\sigma^{\rm fila,gal}_{\rm N,mean} = 2\sqrt{(\sigma^{\rm fila,gal}_{\rm N})^2/N^{\rm cold}_{\rm fila,gal}}\,,
\eeq
in which $(\sigma^{\rm fila,gal}_{\rm N})^2$ is computed by inserting equations (\ref{eq:theta_c_peak}) and (\ref{eq:cir_ap}) into equation (\ref{eq:noise}), and the factor of $2$ results from the fact that the signal is the difference between two parts of the filament. This noise also depends on the host halo mass, which affects the size of the aperture radius.

The expected S/N is shown by the green solid line in Fig.~\ref{fig:ston_fila_gal_wang21}. Although the signal is still too weak to be detected\footnote{We try to measured the kSZ signal from galaxies spinning around filaments by stacking the imprints of the Wang21 catalog galaxies on the Planck 217 GHz map, yet no signal is detected. This is consistent with the theoretical expectation.}, the S/N can be more than one order of magnitude higher than that of the kSZ signal from diffuse gas rotating around filaments, therefore the kSZ signal associated with the rotation of galaxies around filaments may have a better chance to be detected in the future.

In summary, the kSZ signal from diffuse gas rotating around filaments, or galaxies spinning around filaments is very small, at the order of $10^{-3}$ to $10^{-2}~\mu K$. The combination of the filament/galaxy catalog of W21 from SDSS with the CMB temperature map observed by Planck is not expected to show a detection. The major limiting factors are (1) the size of the filament/galaxy sample, this effectively can be translated as the survey volume and the galaxy number density; (2) the size of the beam of the CMB survey, which is important for high-$z$ filaments/galaxies; (3) the detector noise of the CMB survey. Fortunately, upcoming surveys of galaxies and CMB experiments are both improving in those directions, and of course, surveys of the late-time large-scale structure need to overlap with CMB experiments on the sky to be useful for the kSZ measurement. We will focus on estimating what we can achieve with the near-future experiments in the next subsection.

\subsection{Future surveys}
\label{subsec:future}
\begin{table*}
 \centering
 \caption{The specifications of different survey combinations. `Sky coverage overlap' is the overlapping sky area between a galaxy survey and a CMB survey. `Redshift range' is the adopted redshift range of the galaxy survey in our calculations. `$N_{\rm survey}^{\rm fila,cold}/N_{\rm W21}^{\rm fila,cold}$' gives the number of filaments in unit of the number of filaments in W21. `FWHM' is the full width at half-maximum of the beam of CMB experiments. `Detector noise' is the adopted CMB experiment's detector noise. Since DESI mainly observes the northern sky area, we assume a $8000~$deg$^2$ sky area overlap between DESI and future CMB experiments.}
 \begin{tabular}{lccccccccc}
  \hline
  Survey combination & Sky coverage & Redshift range & $N_{\rm survey}^{\rm fila,cold}/N_{\rm W21}^{\rm fila,cold}$ & FWHM & Detector noise & S/N of $\delta T^{\rm fila}_{\rm kSZ}$ & S/N of $\delta T^{\rm fila,gal}_{\rm kSZ}$ \\
  &overlap (deg$^2$)&&&($'$)&($\muk$-$\rm arcmin$)&($R_{\rm fila}=1.8\mpch$)&($M_h=5\times10^{12}\, \rm{M_\odot}/h$) \\
  \hline
  SKA-2+Planck & 30000 & 0-2 & 1868 & 5 & 
  47 & 1.0 & 2.0\\
  SKA-2+SO-LAT & 16500 & 0-2 & 1027 & 1 & 15 & 2.5 & 6.5\\
  SKA-2+CMB-S4 & 30000 & 0-2 & 1868 & 1 & 1 & 3.5 & 16.6\\
  SKA-2+CMB-HD & 20000 & 0-2 & 1245 & 0.25 & 0.5 & 6.7 & 19.7\\
  DESI+Planck & 14000 & 0-1.6 & 20.8 & 5 & 47 & 0.2 & 0.4\\
  DESI+CMB-S4 & 8000 & 0-1.6 & 20.8 & 1 & 1 & 0.3 & 0.9\\
  DESI+CMB-HD & 8000 & 0-1.6 & 20.8 & 0.25 & 0.5 & 0.7 & 1.3\\
  \hline
 \end{tabular}
 \label{tab:survey_detail}
\end{table*}
The estimate for the expected kSZ signal from the Wang21 catalog serves as a benchmark for our forecast for the bigger and deeper surveys. We will now scale up the calculation for future surveys based on the known properties of filaments/galaxies from the previous subsection, and assume perfect overlaps with future CMB experiments. Since we rely on the averaged galaxy redshift to estimate the direction of rotation for each filament, the precision of galaxy redshifts is important. We therefore choose to focus on DESI\footnote{The Dark Energy Spectroscopic Instrument, \url{https://www.desi.lbl.gov}}, SKA-2\footnote{Phase 2 of the Square Kilometre Array, \url{https://www.skao.int}} to represent galaxy spectroscopic redshift surveys, and  SO-LAT\footnote{The Large Aperture Telescope(LAT) of the Simons Observatory, \url{https://simonsobservatory.org}}, CMB-S4\footnote{The next generation Stage-4 ground-based CMB experiment, \url{https://cmb-s4.org}}, CMB-HD\footnote{An Ultra-Deep, High-Resolution Millimeter-Wave Survey Over Half the Sky, \url{https://cmb-hd.org}} to represent the CMB surveys.

\subsubsection{Survey specifications}
DESI is an on-going stage-IV dark energy project covering $14000$ deg$^2$ of the northern sky out to $z=1.7$ \citep{DESI2016}, and SKA-2 will be an ambitious sample variance-limited survey covering $30000$ $\rm deg^2$ of the southern sky at $z<2$ \citep{Yahya2015}. When estimating the $n_{\rm gal}$ of SKA-2, we assume a reference flux sensitivity in Table 4 of \citealt{Yahya2015}. SO-LAT is a southern sky CMB experiment with FWHM$\sim$ $1'$ and $\sim$ 15 $\muk$-$\rm arcmin$ detector noise at the 225GHz frequency band, covering $40\%$ of the sky \citep{Simon2019}. CMB-S4 is a next generation CMB survey with FWHM$\le 1.5'$ and $\sim$1 $\muk$-$\rm arcmin$ detector noise, covering ~$30000$ deg$^2$ of the southern sky \citep{CMBS42019}. CMB-HD is an ultra-high resolution CMB survey with FWHM=$15''$ and  ~0.5 $\muk$-$\rm arcmin$ detector noise, covering ~$20000$ deg$^2$ of the southern sky \citep{CMBHD2020}. The technical specifications of these surveys adopted in our calculation are shown in Table~\ref{tab:survey_detail}. We caution that some of areas for the overlapping sky may be overestimated, but the S/N can be rescaled by the square root of the final areas.

To estimate the expected number density of filaments for a given survey, we use the $n^{\rm cold}_{\rm fila}-n_{\rm gal}$ relation of the Bisous model from the cold sample of Wang21 catalog shown in Fig.~\ref{fig:nfila_ngal_z}, and simply assume that it is redshift-independent. With the galaxy number density $n_{\rm gal}(z)$ of a survey specified, we can obtain the corresponding $n^{\rm cold}_{\rm fila}(z)$. We adopt the averaged number of galaxies per filament from the Wang21 catalog, which is $\approx 9$. This allows us to estimate the number of galaxies in filaments for those surveys. We also adopt the distribution of orientations for filaments from the cold sample of Wang21 catalog (see the bottom panel of Fig. \ref{fig:hist}). We caution that our assumption on the redshift-independence of the $n^{\rm cold}_{\rm fila}-n_{\rm gal}$ relationship is likely to be oversimplified. For example, a deeper galaxy survey will be able to detect more faint galaxies around filaments, but also resolve shorter filaments. In the mean time, the number density of galaxies will evolve with $z$. A more in-depth study on the redshift evolution of $n^{\rm cold}_{\rm fila}-n_{\rm gal}$, including its dependence on the filament finding algorithm and survey selections for galaxies will be needed for realistic forecasts for specific surveys, which is beyond the scope of this paper.

Some details of this procedure are presented in Appendix \ref{append:fila_vs_gal}. Finally, we adopt the same gas density and rotation velocity profiles shown in Fig.~\ref{fig:prof}, interpolating for their redshift evolution.

\begin{figure*}
    \centering
    \includegraphics[width=0.98\columnwidth]{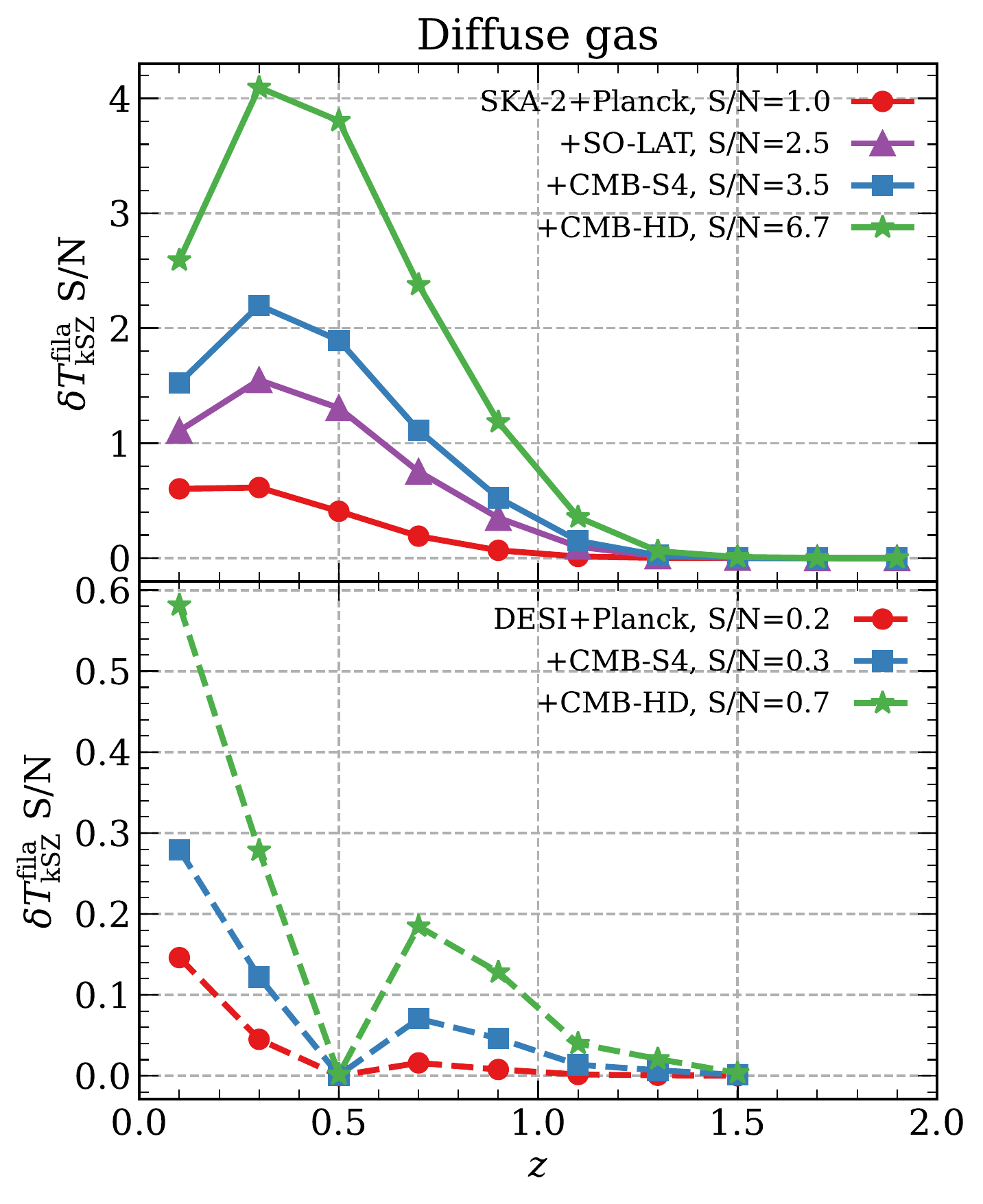}
    \includegraphics[width=\columnwidth]{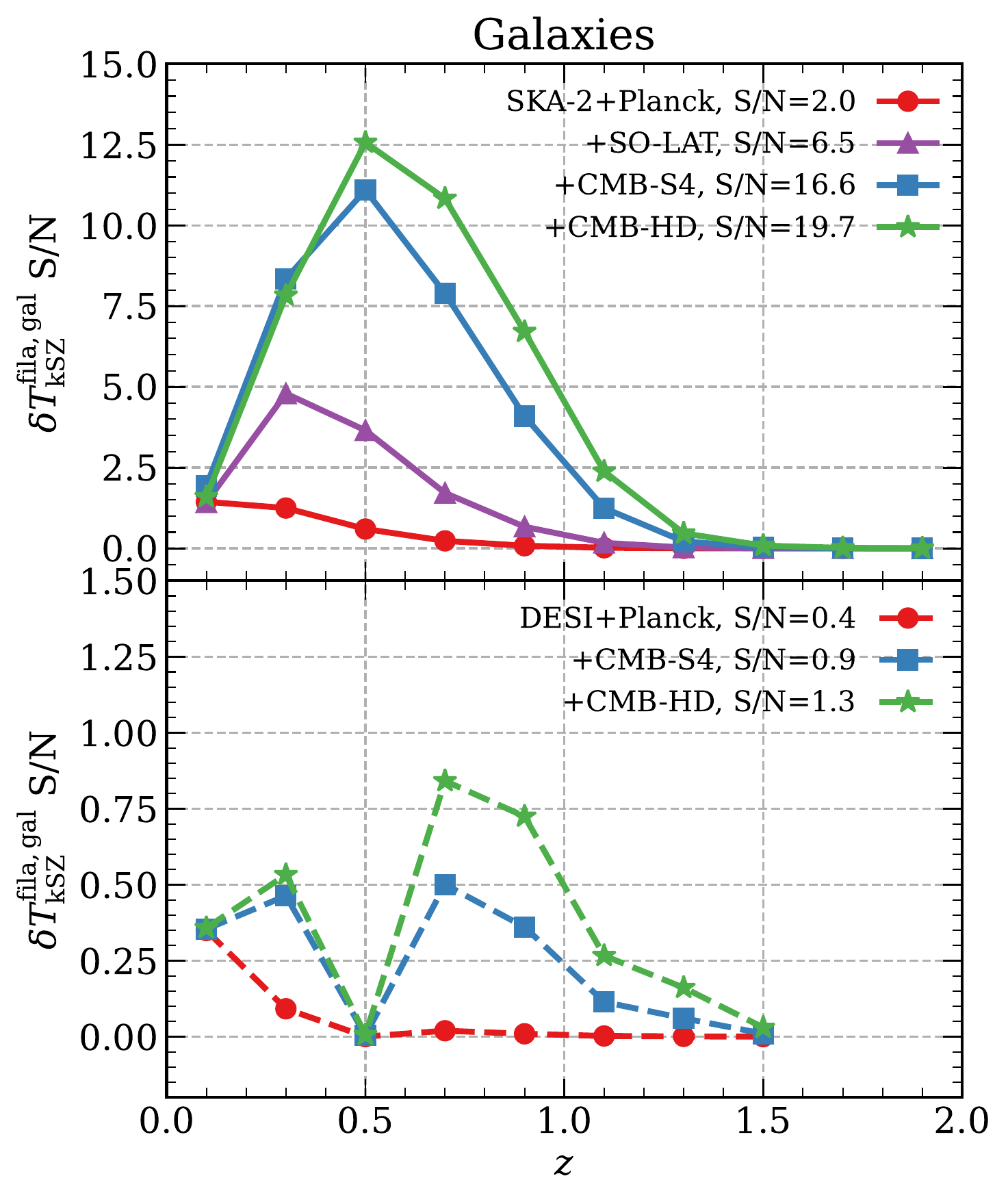}
    \caption{Forecasts for the S/N of the the kSZ dipole signal expected to be observed by different combinations of galaxy redshift surveys and CMB experiments indicated by the legends. The cold filament samples are considered in all evaluations. Left-hand panels show the results for the kSZ dipole induced by the rotation of diffuse filamentary gas; right-hand panels show the cases for the kSZ dipole from galaxies co-rotating with filaments; Top-panels show the combination of SKA-2 with CMB experiments; bottom-panels show the combination of DESI with CMB experiments, all are indicated by the legends. Table~\ref{tab:survey_detail} has the details for the parameters adopted for each survey combination. The total S/N's after summing over all redshift bins for each case are shown in the legend, and also in Table~\ref{tab:survey_detail}. Details about our model assumptions can be found in the text.}
    \label{fig:ston_future}
\end{figure*}
\begin{figure}
    \centering
    \includegraphics[width=\columnwidth]{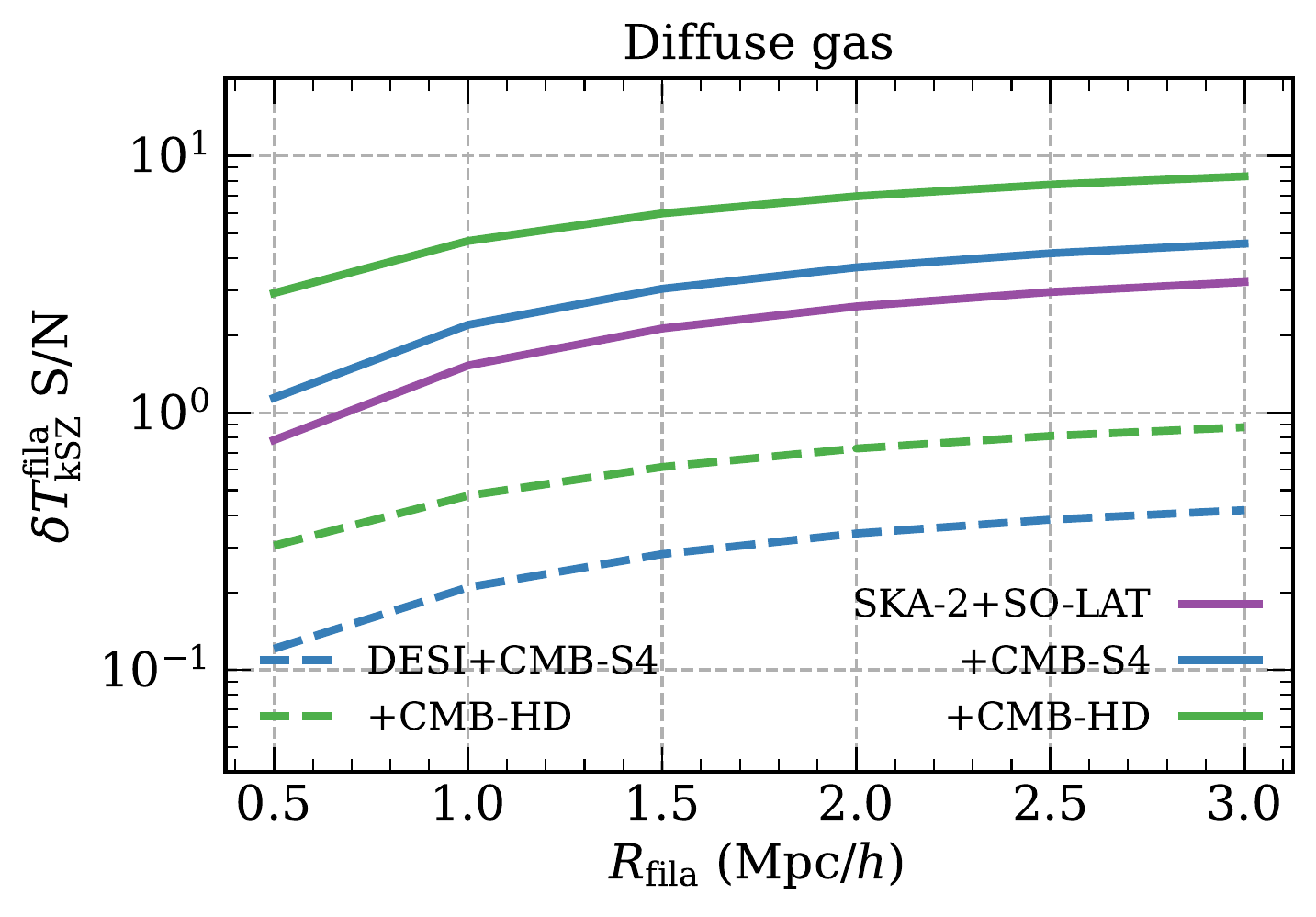}
    \caption{{The total S/N's with all redshifts combined for the kSZ dipole from diffuse filamentary gas $\delta T_{\rm fila}^{\rm cold}$ as a function of filament radius $R_{\rm fila}$. Different survey combinations are indicated in the legend. The total S/N's depends only weakly on $R_{\rm fila}$.}}
    \label{fig:ston_future_fila_Rfila}
\end{figure}
\begin{figure}
    \centering
    \includegraphics[width=\columnwidth]{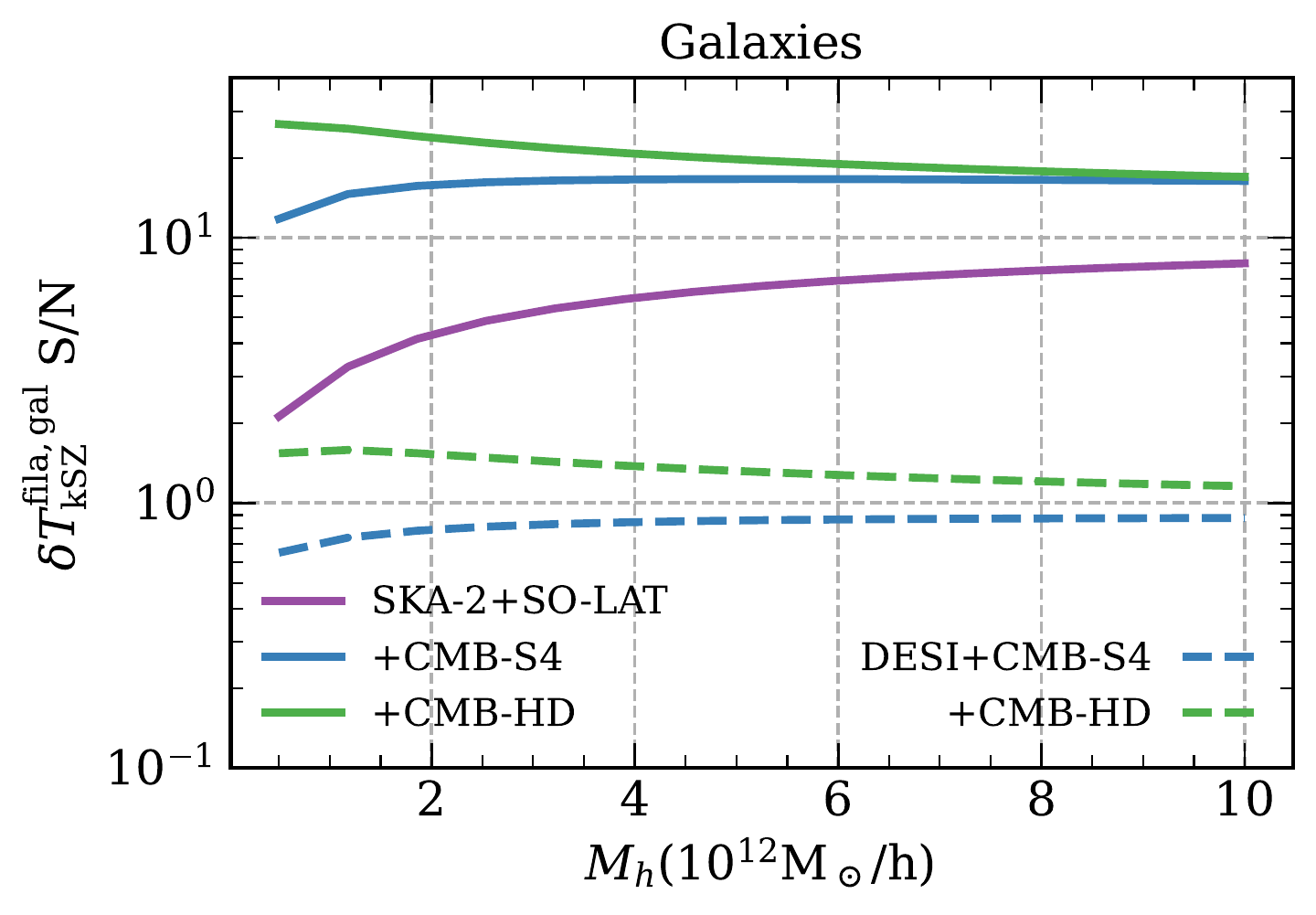}
    \caption{{The total S/N's  with all redshifts combined for the kSZ dipole from galaxies co-rotating with filaments $\delta T_{\rm fila,gal}^{\rm cold}$ as a function of the host halo mass $M_h$. Different survey combinations are indicated in the legend. The total S/N's depends only weakly on $M_h$.}}
    \label{fig:ston_future_galfila_Mh}
\end{figure}

\subsubsection{S/N predictions}

Fig.~\ref{fig:ston_future} presents our forecasts for the kSZ signal around filaments by all gas (left), and the kSZ signal by galaxies rotating around filaments (right). We see the clear dependence of the S/N on the number of filaments for the same CMB experiment. The catalog size $N$ influences the statistical error of the detection by $\propto 1/\sqrt{N}$, as illustrated by equations (\ref{eq:err_fila_per}) and (\ref{eq:err_filagal}). Therefore, the S/N's of the DESI-based combinations are much lower than those of SKA-2-based ones, with the latter typically having a factor of $\sim$50 more filaments. It is also worth noting that the redshifts where S/N curves peak are usually higher than that of peaks of the redshift distribution of the galaxy sample. This is mainly due to the smaller angular size of the filter at high-$z$, which helps to suppress more large-scale CMB fluctuations compared to the low-$z$ cases.

Beside sample variance, the next dominant factor for the S/N appears to be the beam of CMB experiments, with a clear trend that the S/N increases with a decreasing FWHM of the beam. Next, with the same beam, especially when it is relatively small, the detector noise starts to become important. For example, the survey combinations of `SKA-2+CMB-S4' and `SKA-2+SO-LAT' have the same FWHM but different detector noises. The former filament catalog is $1868/1028\approx1.8$ times of that of the latter, resulting in a $\approx 1.3$ times of S/N if it is sample variance dominated. By comparing the blue and purple lines in Fig. \ref{fig:ston_future}, the final S/N differences of these two combinations are $1.46$ times for $\delta T_{\rm kSZ}^{\rm fila}$ and $3.11$ times for $\delta T_{\rm kSZ}^{\rm fila,gal}$. This suggests that the difference of the detector noise is playing a role here, and that the kSZ signal associated with the rotation of galaxies is more sensitive to it than that of the rotating gas around filaments. This is expected as the size of the compensated-top-hat filter for the former is usually smaller, thus being more effective in suppressing large-scale CMB fluctuations.

For the kSZ signal from gas rotating around filaments, we also investigate the dependence of the total S/N's on our assumption about the radius of filament $R_{\rm fila}$. Fig.~\ref{fig:ston_future_fila_Rfila} presents the total S/N v.s. $R_{\rm fila}$. As expected, thicker filaments have higher S/N's, but the dependence is weak.

For the kSZ signal from galaxies rotating around filaments, we also test the dependence of the total S/N's on our assumption about the galaxy's host halo mass $M_h$, shown in Fig. \ref{fig:ston_future_galfila_Mh}. We again see a very weak dependence on $M_h$ for all cases. Therefore, the predictions is stable against the choice for $M_h$. Besides, we also see a flattish scaling pattern of S/N-$M_h$ relation at the large $M_h$ end. This trend can be understood from Fig.~\ref{fig:ston_fila_gal_wang21}, in that both the signal and noise after applying the AP filter increase and cancel each other towards a larger $M_h$.

Table~\ref{tab:survey_detail} summarises the details of survey combinations together with the total expected S/N for each combination. Limited mainly by the overlapping survey area, we find that surveys like DESI are unlikely to give a filament-rotation kSZ signal. A SKA-2-like survey plus a next stage CMB experiment is needed to have a detection of kSZ dipole from the diffuse gas at a few and up to ten $\sigma$'s level. For galaxies rotating around filaments, the expected S/N can be larger than 10 $\sigma$. 

\section{Conclusion and discussion}
\label{sec:conclusion}

We have developed a theoretical framework for studying the kSZ signal induced by the spin of intergalactic filaments. 
The characteristic feature of this signal is a kSZ temperature dipole imprinted on the CMB, which we call $\delta T^{\rm fila}_{\rm kSZ}$. In addition, the gas halos of galaxies embedded in filaments are expected to rotate together with the filaments, they too induce kSZ temperature dipoles associated with the gas halos, which we call $\delta T^{\rm fila,gal}_{\rm kSZ}$. 

A detection of $\delta T^{\rm fila}_{\rm kSZ}$ and $\delta T^{\rm fila,gal}_{\rm kSZ}$ requires identifying filaments in the cosmic web from a galaxy redshift survey. Because of the weak signal, a detection may also require estimates from redshift-space filament shapes (as in W21) of filament rotation directions. This would make the kSZ measurement correlated somewhat with the galaxy redshift-space measurement, but not completely, since kSZ probes velocities within filaments, independent of their real-space shapes. In addition, a CMB experiment overlapping with the galaxy redshift survey on the same sky is needed for the kSZ measurements. We based our analysis on the properties of filaments/galaxies of the Wang21 catalog from SDSS and their combination with CMB temperature maps from Planck. We have also verified our calculations for the noise using observed CMB temperature maps from Planck.

We find the amplitude of the kSZ dipole from the diffuse gas to be small, $\delta T^{\rm fila}_{\rm kSZ} \sim O(10^{-3})\mu K$, but the kSZ signal from gas halos $\delta T^{\rm fila,gal}_{\rm kSZ}$ can be one order of magnitude larger, $\delta T^{\rm fila,gal}_{\rm kSZ} \sim  O(10^{-2})\mu K$. This is mainly due to the much higher gas density in gas halos. We also find that the S/N of the kSZ dipole from the combination of SDSS+Planck to be very small, and therefore is currently not detectable. The major limiting factor is the size of the filament sample.

We then apply similar analyses to predict the detectability of the kSZ dipole for the combinations of future galaxy redshift surveys and next-stage CMB experiments. We consider DESI and SKA-2 as our galaxy survey representatives, and SO-LAT, CMB-S4, CMB-HD as the target CMB experiments. Our calculations show that, the detection S/N's of these future surveys for $\delta T^{\rm fila}_{\rm kSZ}$ are around $O(0.1)$ to $O(1)$, while those of $\delta T^{\rm fila,gal}_{\rm kSZ}$ range from $O(1)$ to $O(10)$. In particular, we find that DESI-like surveys are unlikely to give a filament-rotation kSZ signal. A SKA-2-like survey plus a next stage CMB experiment is needed to have a detection of kSZ dipole from the diffuse gas at a few and up to ten $\sigma$'s level. For galaxies rotating around filaments, the expected S/N can be larger than 10 $\sigma$. 

A detection for the kSZ dipole signal can help to establish firm evidence for the spin of filaments. It may also provide constraints for the baryon content in filaments. As long as we have some prior knowledge about the filament rotational velocity profile, it can be deconvolved from the measured $\delta T^{\rm fila}_{\rm kSZ}(d)$ and unveil the information of the baryon density profile. Since kSZ signal does not depend on the gas temperature, it is an important complement for other methods of detecting the missing baryons within filaments. Moreover, the spin of a filament should induce a lensing dipole of the same orientation as the kSZ one through the gravitomagnetic effect. The cross-correlations between these two redshift dipoles may provide a way to detect the gravitomagnetic effect on cosmological scales \citep{Schafer2006,Cristian2022}.

Limited by our knowledge about the redshift evolution of filaments and galaxy properties in filaments, and their number density redshift distribution associated with a specific filament detection method and a certain kind of tracer, we adopt a series of simplifications in our calculations. The major ones include (1) a rotational velocity profile taken from a dark matter simulation measurement (X21) and its redshfit dependence to be equal to that of the voticity power spectrum; (2) the redshift-independent $n_{\rm fila}^{\rm cold}-n_{\rm gal}$ relation, cold filament orientation distribution, the filament radius $R_{\rm fila}$, and the galaxy number per cold filament, all taken empirically from a low redshift filament catalog constructed by the Bisous model \citep{Tempel2014} regardless of the target tracer type; (3) a Gaussian projected gas profile within the host halos, while other profiles can be adopted (e.g.  \citealt{Cavaliere1976,Wu2020}); (4) a universal gas fraction $f_{\rm gas}$ independent of the host halo mass~\citep{Lim2020}, even though there is still no good consensus for $f_{\rm gas}$ predicted from different hydro-simulations \citep[e.g.][]{Oppenheimer2021}. 

The minor assumptions are that (1) we use the mean length of filaments to represent the measured length of filaments in our calculations, (2) we ignore the correlation between different $\theta_Y$ bins in the S/N estimation of $\delta T_{\rm kSZ}^{\rm fila}$, (3) we adopt a constant halo optical depth instead of a $\tauT$ distribution in the S/N estimation of $\delta T^{\rm fila,gal}_{\rm kSZ}$. 

Despite these assumptions, we have demonstrated where possible that some of 
the uncertainties on the final S/N estimation should be minor, as discussed in the other parts of this paper. We caution that our final S/N estimations are more meaningful in their order of magnitude. Being uncertain in details though, the theoretical framework we develop in this paper itself has captured the major features describing the filament spin induced kSZ signals. In the future, we will resort to dark matter and hydro simulations to pin down some of the above uncertainties and make a more accurate S/N prediction based on a realistic mock catalog of filaments associated with specific methods for filament detection. 

There are some other issues whose influences we do not discuss in detail in this paper. For example, with the help of a realistic mock filament catalog from simulations, we will be also able to study a possible systematic error in the kSZ signal detection due to possible mis-identifications of filament spin directions, which are assumed from the averaged redshift of galaxies around each filament. CMB foregrounds like tSZ, CIB, and galaxy emission are a worry for the detection, but in principle they can be removed because of their frequency dependence in different CMB bands. Most importantly, they should not have a dipole pattern, as our kSZ estimator picks up. But it remains to be proven how well this contamination can be removed practically.

Finally, we notice that the Stage IV surveys we consider here have not exhausted all the available cosmological information in our universe. In particular the Stage IV galaxy surveys still have limited filament numbers at high redshifts due to the observational magnitude limits. A more ambitious and ideal survey combination can in principle further improve the S/N, e.g. the Stage V surveys \citep{Dodelson2016,Schlegel2022}.

\section*{Acknowledgements}

We thank Junde Chen, Zhiqi Huang, Elmo Tempel and Zhejie Ding for useful discussions. YZ acknowledges the support from the National Natural Science Foundation of China (NFSC) through grant 12203107 and the support from the Guangdong Basic and Applied Basic Research Foundation with No.2019A1515111098, the science research grants from the China Manned Space Project with NO.CMS-CSST-2021-A02. YC acknowledges the support of the Royal Society through the award of a University Research Fellowship and an Enhancement Award. WSZ acknowledges the support from the National Natural Science Foundation of China (NFSC) through grants 11733010 and 12173102. MN acknowledges support by the Spanish grant PID2020-114035GB-100 (MINECO/AEI/FEDER, UE). 

For the purpose of open access, the authors have applied a CC BY public copyright licence to any Author Accepted Manuscript version arising.

\section*{Data Availability}
The data underlying this article are available on request to the corresponding author.


\bibliographystyle{mnras}
\bibliography{mybib} 


\appendix
\section{Analytical expressions of the kSZ signal profile from the spin of filaments}
\label{append:analytical_kSZ_profile}
The expected kSZ temperature profile from diffuse gas induced by the rotation of filaments can be calculated with equation (\ref{eq:ksz_profile}). We will lay down its analytical expression here. The expression for the integration of equation (\ref{eq:ksz_profile}) depends on the relative differences between $R_{\rm fila}$, $2\mpch$ and $R_{\rm lim}$. $R_{\rm fila}$ is the baryonic radius of a filament, which indicates the boundary between a filament and a wall or void outside it. It characterizes the baryon density profile described in equation~(\ref{eq:den_prof}). $2\mpch$ is the transition scale in the baryon rotational velocity profile in equation~(\ref{eq:vel_prof}). $R_{\rm lim}$ is the boundary of integration for the filament, up to which the kSZ signal contribution of baryons in a filament is considered in our evaluation. As generally specified in our study, we focus on the case with $R_{\rm fila} \le R_{\rm lim}$. In the case of $R_{\rm fila}\le R_{\rm lim} < 2\mpch$, equation (\ref{eq:ksz_profile}) can be derived as
\begin{widetext}

\begin{eqnarray*}
\delta T_{\ksz}(d)&=& \frac{T_0\sigma_{\rm T}}{a^2c}\int_{\theta_f}^{\theta_b}{\rm d}\theta \frac{d}{\cos\theta} n_e(\frac{d}{\cos\theta})v_{\rm rot}(\frac{d}{\cos\theta})=\frac{T_0\sigma_{\rm T}}{a^2c}\alpha_{\rm curl}n_{e,0}d \\
&\times&\left\{
\begin{array}{cc}
\left[\frac{170}{\sqrt{1+(d/r_c)^2}}\tanh^{-1}\frac{\sqrt{1-(d/R_{\rm fila})^2}}{\sqrt{1+(d/r_c)^2}} + \frac{632570}{13721} \ln\frac{R_{\rm lim}+\sqrt{R_{\rm lim}^2-d^2}}{R_{\rm fila}+\sqrt{R_{\rm fila}^2-d^2}}\right]\,,& |d|< R_{\rm fila} \\
\quad\\
\left[\frac{632570}{13721} \ln\frac{R_{\rm lim}+\sqrt{R_{\rm lim}^2-d^2}}{|d|}\right] \,. & R_{\rm fila} \le |d| \le R_{\rm lim}
\end{array}
	\right.
\label{eq:kSZ_profile_analytical1}
\end{eqnarray*}
\end{widetext}
In the case of $R_{\rm fila} \le 2\mpch \le R_{\rm lim}$, equation (\ref{eq:ksz_profile}) can be derived as
\begin{widetext}
\begin{eqnarray*}
\delta T_{\ksz}(d)&=& \frac{T_0\sigma_{\rm T}}{a^2c}\int_{\theta_f}^{\theta_b}{\rm d}\theta \frac{d}{\cos\theta} n_e(\frac{d}{\cos\theta})v_{\rm rot}(\frac{d}{\cos\theta})=\frac{T_0\sigma_{\rm T}}{a^2c}\alpha_{\rm curl}n_{e,0}d \\
&\times&\left\{
\begin{array}{cc}
\left[\frac{170}{\sqrt{1+(d/r_c)^2}}\tanh^{-1}\frac{\sqrt{1-(d/R_{\rm fila})^2}}{\sqrt{1+(d/r_c)^2}} + \frac{632570}{13721} \ln\frac{2+\sqrt{4-d^2}}{R_{\rm fila}+\sqrt{R_{\rm fila}^2-d^2}}\right.\\
\left.+ \frac{1004670}{13721} \ln\frac{R_{\rm lim}+\sqrt{R_{\rm lim}^2-d^2}}{2+\sqrt{4-d^2}}-\frac{186050}{13721}\left(\sqrt{R_{\rm lim}^2-d^2}-\sqrt{4-d^2}\right)\right]\,,& |d|<R_{\rm fila} \\
\quad\\
\left[\frac{632570}{13721} \ln\frac{2+\sqrt{4-d^2}}{|d|}+ \frac{1004670}{13721} \ln\frac{R_{\rm lim}+\sqrt{R_{\rm lim}^2-d^2}}{2+\sqrt{4-d^2}}-\frac{186050}{13721}\left(\sqrt{R_{\rm lim}^2-d^2}-\sqrt{4-d^2}\right)\right] \,, & R_{\rm fila}\le |d| < 2\mpch \\
\quad\\
\left[\frac{1004670}{13721} \ln\frac{R_{\rm lim}+\sqrt{R_{\rm lim}^2-d^2}}{|d|}-\frac{186050}{13721}\sqrt{R_{\rm lim}^2-d^2}\right] \,. & 2\mpch\le |d| \le R_{\rm lim}
\end{array}
	\right.
\label{eq:kSZ_profile_analytical2}
\end{eqnarray*}
\end{widetext}
In the case of $2\mpch < R_{\rm fila} \le R_{\rm lim}$, equation (\ref{eq:ksz_profile}) can be derived as
\begin{widetext}
\begin{eqnarray*}
\delta T_{\ksz}(d)&=& \frac{T_0\sigma_{\rm T}}{a^2c}\int_{\theta_f}^{\theta_b}{\rm d}\theta \frac{d}{\cos\theta} n_e(\frac{d}{\cos\theta})v_{\rm rot}(\frac{d}{\cos\theta})=\frac{T_0\sigma_{\rm T}}{a^2c}\alpha_{\rm curl}n_{e,0}d \\
&\times&\left\{
\begin{array}{cc}
\left[\frac{170}{\sqrt{1+(d/r_c)^2}}\tanh^{-1}\frac{\sqrt{1-(d/2)^2}}{\sqrt{1+(d/r_c)^2}}+\frac{270}{\sqrt{1+(d/r_c)^2}}\left(\tanh^{-1}\frac{\sqrt{1-(d/R_{\rm fila})^2}}{\sqrt{1+(d/r_c)^2}}-\tanh^{-1}\frac{\sqrt{1-(d/2)^2}}{\sqrt{1+(d/r_c)^2}}\right)\right. \\
\left.-\frac{50r_c}{\sqrt{1+(d/r_c)^2}}\left(\tan^{-1}\frac{\sqrt{(R_{\rm fila}/d)^2-1}}{\sqrt{1+(r_c/d)^2}}-\tan^{-1}\frac{\sqrt{(2/d)^2-1}}{\sqrt{1+(r_c/d)^2}}\right)\right.\\
\left.+\frac{1004670}{13721}\ln\frac{R_{\rm lim}+\sqrt{R_{\rm lim}^2-d^2}}{R_{\rm fila}+\sqrt{R_{\rm fila}^2-d^2}}-\frac{186050}{13721}\left(\sqrt{R_{\rm lim}^2-d^2}-\sqrt{R_{\rm fila}^2-d^2}\right)\right]\,,& |d|<2\mpch \\
\quad\\
\left[\frac{270}{\sqrt{1+(d/r_c)^2}}\tanh^{-1}\frac{\sqrt{1-(d/R_{\rm fila})^2}}{\sqrt{1+(d/r_c)^2}}-\frac{50r_c}{\sqrt{1+(d/r_c)^2}}\tan^{-1}\frac{\sqrt{(R_{\rm fila}/d)^2-1}}{\sqrt{1+(r_c/d)^2}}\right.\\
\left.+\frac{1004670}{13721}\ln\frac{R_{\rm lim}+\sqrt{R_{\rm lim}^2-d^2}}{R_{\rm fila}+\sqrt{R_{\rm fila}^2-d^2}}-\frac{186050}{13721}\left(\sqrt{R_{\rm lim}^2-d^2}-\sqrt{R_{\rm fila}^2-d^2}\right)\right]\,, & 2\mpch\le |d| < R_{\rm fila} \\
\quad\\
\left[\frac{1004670}{13721} \ln\frac{R_{\rm lim}+\sqrt{R_{\rm lim}^2-d^2}}{|d|}-\frac{186050}{13721}\sqrt{R_{\rm lim}^2-d^2}\right] \,. & R_{\rm fila}\le |d| \le R_{\rm lim}
\end{array}
	\right.
\label{eq:kSZ_profile_analytical3}
\end{eqnarray*}
\end{widetext}

The above analytical expressions have been checked against numerical integration, and they yield the same answers.

\subsection{Convergence tests for $R_{\rm lim}$}
\label{append:R_lim_convergence}
\begin{figure}
    \centering
    \includegraphics[width=\columnwidth]{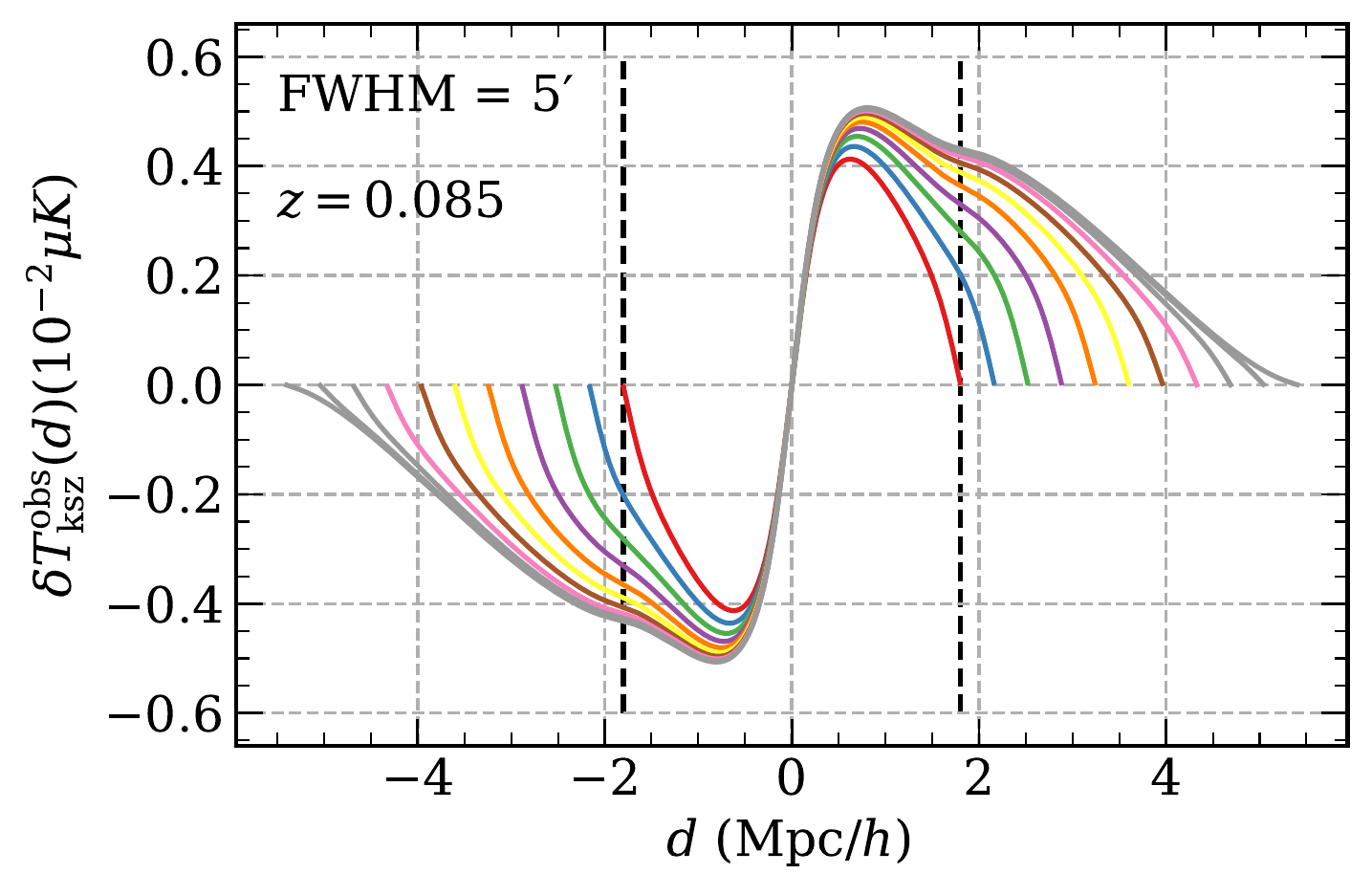}
    \caption{$\delta T^{\rm obs}_{\rm kSZ}$ calculated with equations (\ref{eq:T_ksz_obs}) and (\ref{eq:ksz_profile}) with different upper bound of integration $R_{\rm lim}$. The filament is assumed to be perpendicular to the LOS. From the inside to the outside, the curves represents cases with $R_{\rm lim}=[1.0,1.2,1.4,1.6,1.8,2.0,2.2,2.4,2.6,2.8,3.0]R_{\rm fila}$ and $R_{\rm fila} = 1.8\mpch$. The choices of parameters in the calculation correspond to the Wang21 catalog in combination with Planck.}
    \label{fig:ksz_profile_Rlim}
\end{figure}
\begin{figure}
    \centering
    \includegraphics[width=\columnwidth]{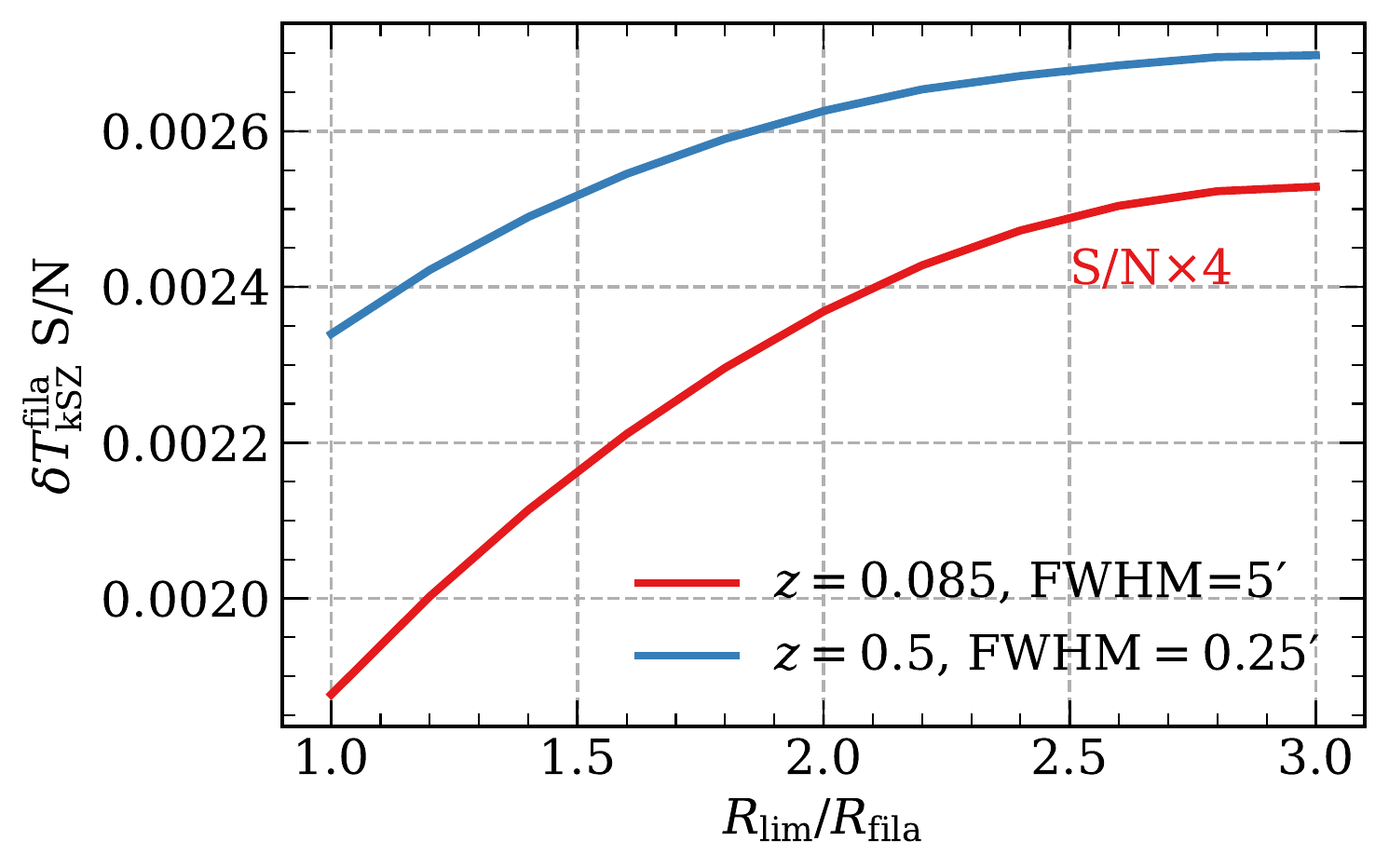}
    \caption{Similar to Fig.~\ref{fig:ksz_profile_Rlim} but showing the S/N of $\delta T^{\rm fila}_{\rm kSZ}$ of different $R_{\rm lim}$. The two cases of filaments at different redshifts and with different FWHMs for the CMB experiments are show in blue and red colours. The red line has been multiplied by a factor of four for a better illustration.}
    \label{fig:ston_Rlim}
\end{figure}
In this subsection, we vary $R_{\rm lim}$ and check the convergence of the kSZ signal from the spin of a filament. As the upper bound of our integration increases, we expect more gas/galaxies will contribute to the kSZ signal, but because both the gas density and rotational velocity profiles are decreasing, we anticipate the contribution at large radii to be small.

Fig.~\ref{fig:ksz_profile_Rlim} shows the kSZ signal profile of a filament with $R_{\rm fila}=1.8\mpch$ at $z=0.085$, convolved with a FWHM$=5'$ beam. As $R_{\rm lim}$ increases, there are more baryons involved in the integration and the signal increases. Meanwhile, the signal can be observed in a wider $d$ range, although the signal at larger $d$ is small due to shorter integration length and decreasing density and rotation velocity. These result in a higher S/N of $\delta T^{\rm fila}_{\rm kSZ}$ for a larger $R_{\rm lim}$, as shown in Fig. \ref{fig:ston_Rlim}.

In Fig. \ref{fig:ston_Rlim}, the red line shows the $\delta T^{\rm fila}_{\rm kSZ}$ S/N's, computed by equation (\ref{eq:ston_fila_cold_per}), for different $R_{\rm lim}$ and for the setup of Wang21 catalog and Planck. The S/N increases and converges towards the value at $R_{\rm lim}=3R_{\rm fila}$. The convergence is expected since both the density and rotational velocity profiles decrease with radius. Another case at $z=0.5$ and FWHM$=0.25'$ is shown in blue for comparison. A similar trend is observed.

Based on the above two figures, we draw our default choice for the upper limit of our integration at $R_{\rm lim}=2 R_{\rm fila}$, which should have more than 90$\%$ of the total S/N.

\section{Wang21 catalog details}
\label{append:wang21_catalog}
\begin{figure}
    \centering
    \includegraphics[width=\columnwidth]{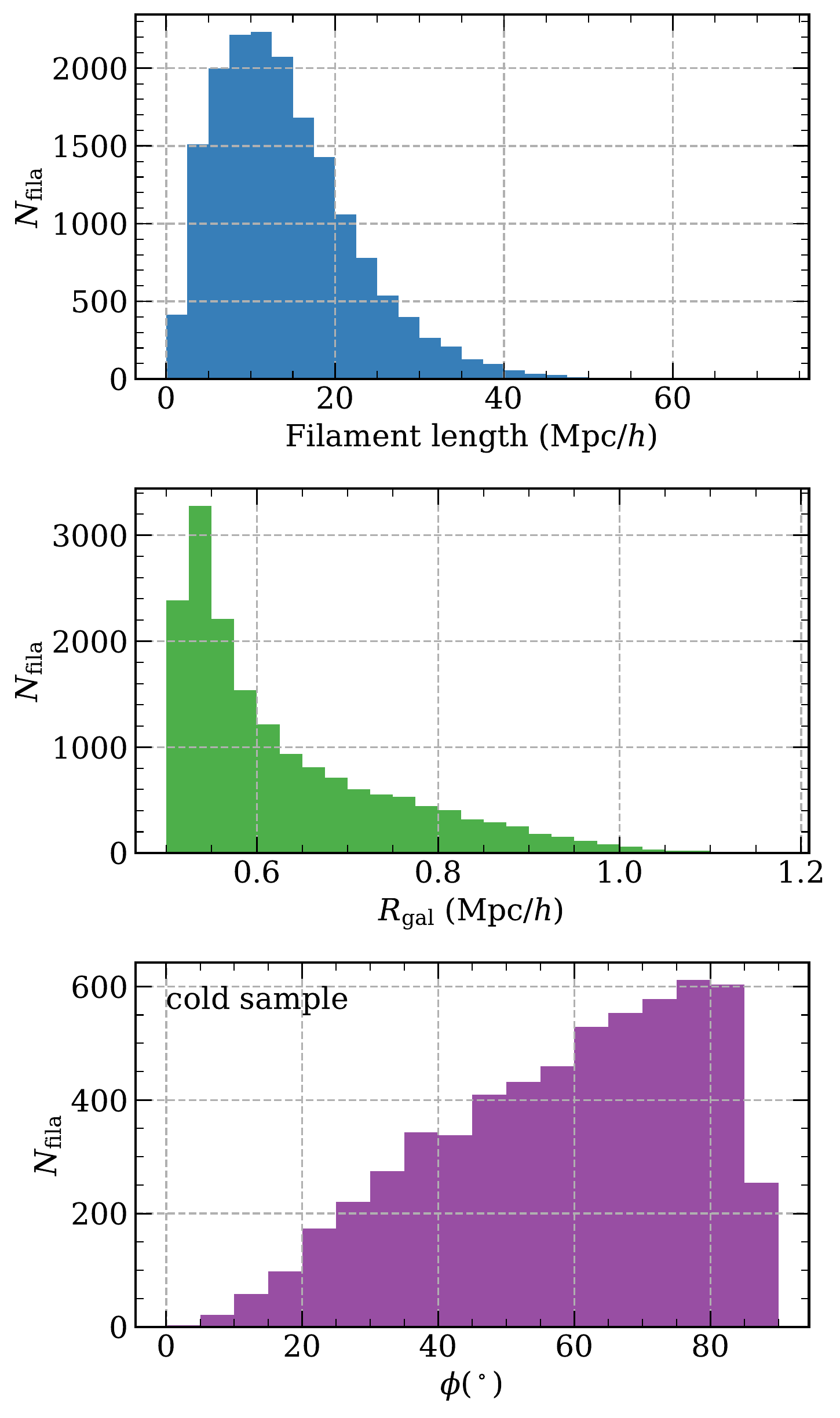}
    \caption{\textit{Top and middle panels}: The histograms of filament length and galaxy radius distribution in the Wang21 catalog. The catalog is selected to have at least 6 galaxies in each filament, 3 on each side. There are in total 17181 filaments. \textit{Bottom panel}: the filament orientation distribution of the sub-sample of the dynamically cold filaments. $\phi$ is the angle between LOS and filament orientation, and $\phi = 90^\circ$ means a filament is perpendicular to the LOS.}
    \label{fig:hist}
\end{figure}

Filaments in the Wang21 catalog are constructed by the Bisous model \citep{Tempel2014}, a marked object point process with interactions, from the Legacy survey of SDSS DR12 data \citep{SDSSdr12}. Discarding galaxies near the edge of survey area, the remaining sky coverage of survey is 7221~deg$^2$. The filaments stay in a redshift range $0<z<0.2$, with the mean redshift $z_{\rm avg} = 0.083$. Under the requirement that there are at least 6 galaxies within one filament, 3 on each side, the final size of Wang21 catalog is 17181 filaments. These filaments have a minimum $L_{\rm min}=1.43\mpch$, a maximum $L_{\rm max} = 70.5\mpch$, the mean $L_{\rm avg} = 14.1\mpch$ and a median length $L_{\rm med} = 12.7\mpch$. The radius of a filament is defined by its galaxy distribution, denoted by the `galaxy radius' $R_{\rm gal}$ in this work. By construction, the filaments have a coherent cylinder radius size chosen by the filament searcher, which is $0.5\mpch$ in the Wang21 catalog \citep{Tempel2014}. In turn, $R_{\rm gal}^{\rm min} = 0.50\mpch$, $R_{\rm gal}^{\rm max} = 1.17\mpch$, $R_{\rm gal}^{\rm avg} = 0.63\mpch$ and $R_{\rm gal}^{\rm med} = 0.59\mpch$. The histograms of filament length and radius distribution are shown in the top and middle panels of Fig. \ref{fig:hist}. Motivated by the $R_{\rm gal}$ distribution of the Wang21 catalog, we assume that all filament galaxies reside in regions with $r<2\mpch$ in our calculation.

In order to maximise the significance of their detection, W21 selected a `dynamically cold' sample from the their catalog. The `coldness' of a filament is defined by $z_{\rm rms}/\Delta z_{\rm AB}$, in which $z_{\rm rms}$ is the root mean square of the galaxy redshifts inside this filament, and the dynamically cold filaments are those with $z_{\rm rms}/\Delta z_{\rm AB} < 1$. This sub-sample consists of of $N^{\rm cold}_{\rm fila} = 5964$ filaments which orientates within $0^\circ \le \phi \le 90^\circ$, with a mean redshift $z_{\rm avg}^{\rm cold} = 0.085$. Although the orientations of filaments are not specifically selected in this sub-sample, the coldness of filaments are correlated with the filament orientations (Fig. 2 of W21), therefore the orientation of filaments is not randomly distributed in this sub-sample, shown in the bottom panel of Fig. \ref{fig:hist}. After discarding the potential interlopers, there remains $N^{\rm cold}_{\rm fila,gal}=54417$ filament member galaxies in this cold sub-sample, which reside within a distance of $1\mpch$ to the filament spine.

Furthermore, we specify a `cold $\perp$ sample', consisting of those filaments lying perpendicular to the LOS, namely the filaments with $\phi\approx90^\circ$. These filaments are less affected by the baryons' helical movement along the filament spine, which may causes a systematic error in the filament spin measurement. If we relax the criteria of being perpendicular to $\phi>70^\circ$, there are $N^{\rm cold,\perp}_{\rm fila} = 2048$ such filaments in the Wang21 catalog. The mean redshift of this sub-sample is $z_{\rm avg}^{\rm cold,\perp}=0.085$.

\section{Filament number density vs galaxy number density}
\label{append:fila_vs_gal}
\begin{figure}
    \centering
    \includegraphics[width=\columnwidth]{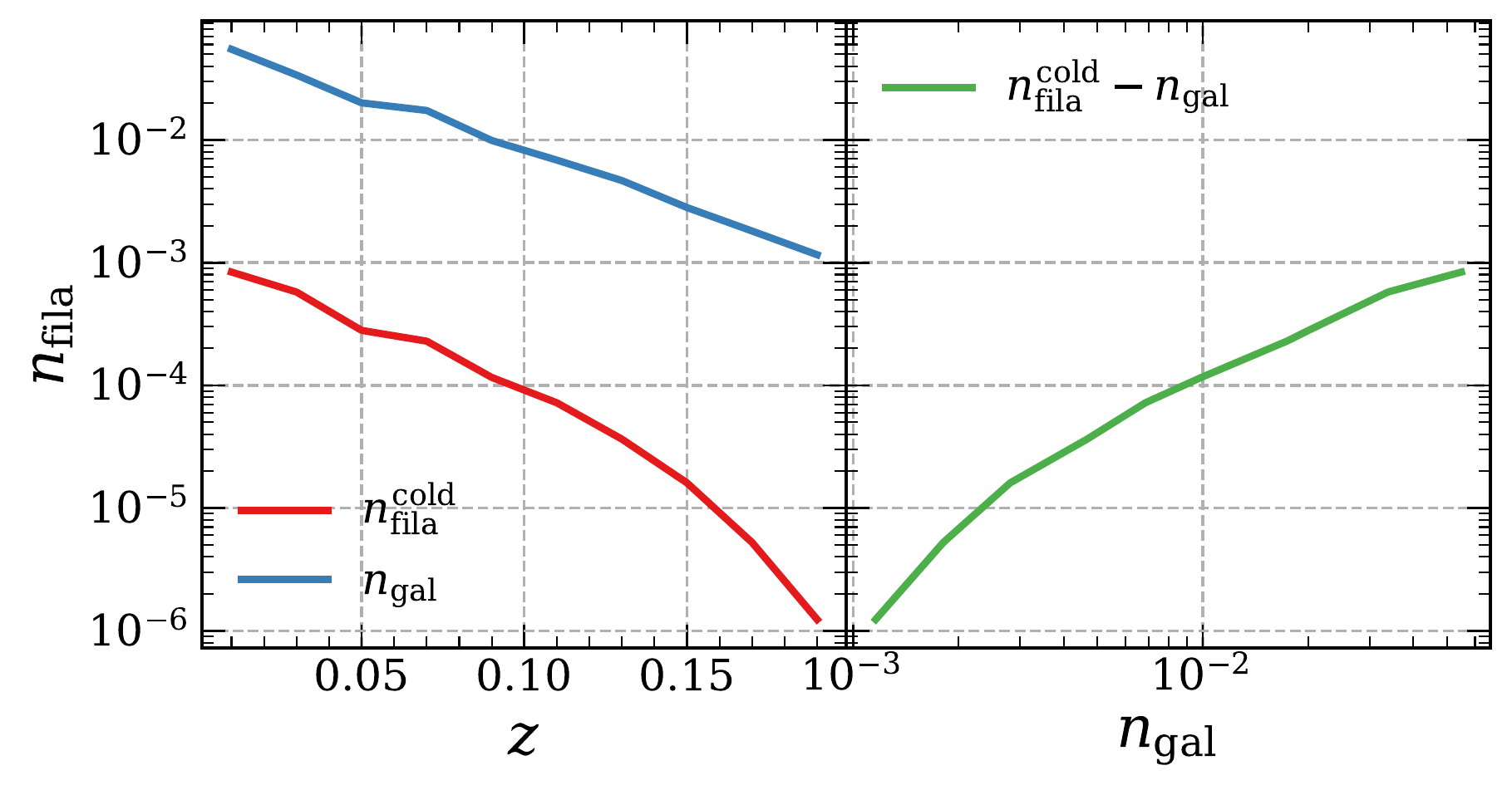}
    \caption{\textit{Left}: The redshift distributions of $n_{\rm fila}^{\rm cold}$ from the  Wang21 catalog, and $n_{\rm gal}$ of all SDSS galaxies at low redshifts, where the redshift range $0<z<0.2$ is divided into 10 bins. \textit{Right}: The the observed relation between the number density of filaments versus the number density of galaxies, $n_{\rm fila}^{\rm cold}-n_{\rm gal}$.}
    \label{fig:nfila_ngal_z}
\end{figure}
In this section we quantify the empirical $n_{\rm fila}-n_{\rm gal}$ relation from the Wang21 catalog and apply it to predict the corresponding $n_{\rm fila}(z)$ for future surveys. The galaxy and filament number densities we mention in this paper are comoving number densities.

In the left panel of Fig. \ref{fig:nfila_ngal_z}, we present the number densities of filaments from the cold sample (red line) and the number densities of all SDSS galaxies (blue line) versus redshift. The number densities of filaments is plotted against the number density of galaxies from the SDSS main galaxy sample+LOWZ sample on the right (green line). 

In general, the filament number densities evolve coherently with that of galaxies, indicating that $n_{\rm gal}$ is the major factor that determines the observed number density of filaments. The smoothness of the relation justify using it to estimate the number density of filaments based on our knowledge of $n_{\rm gal}$.

\subsection{Number density of filaments and galaxies in filaments from DESI and SKA-2}
\label{append:obtain_nfila}
\begin{figure}
    \centering
    \includegraphics[width=\columnwidth]{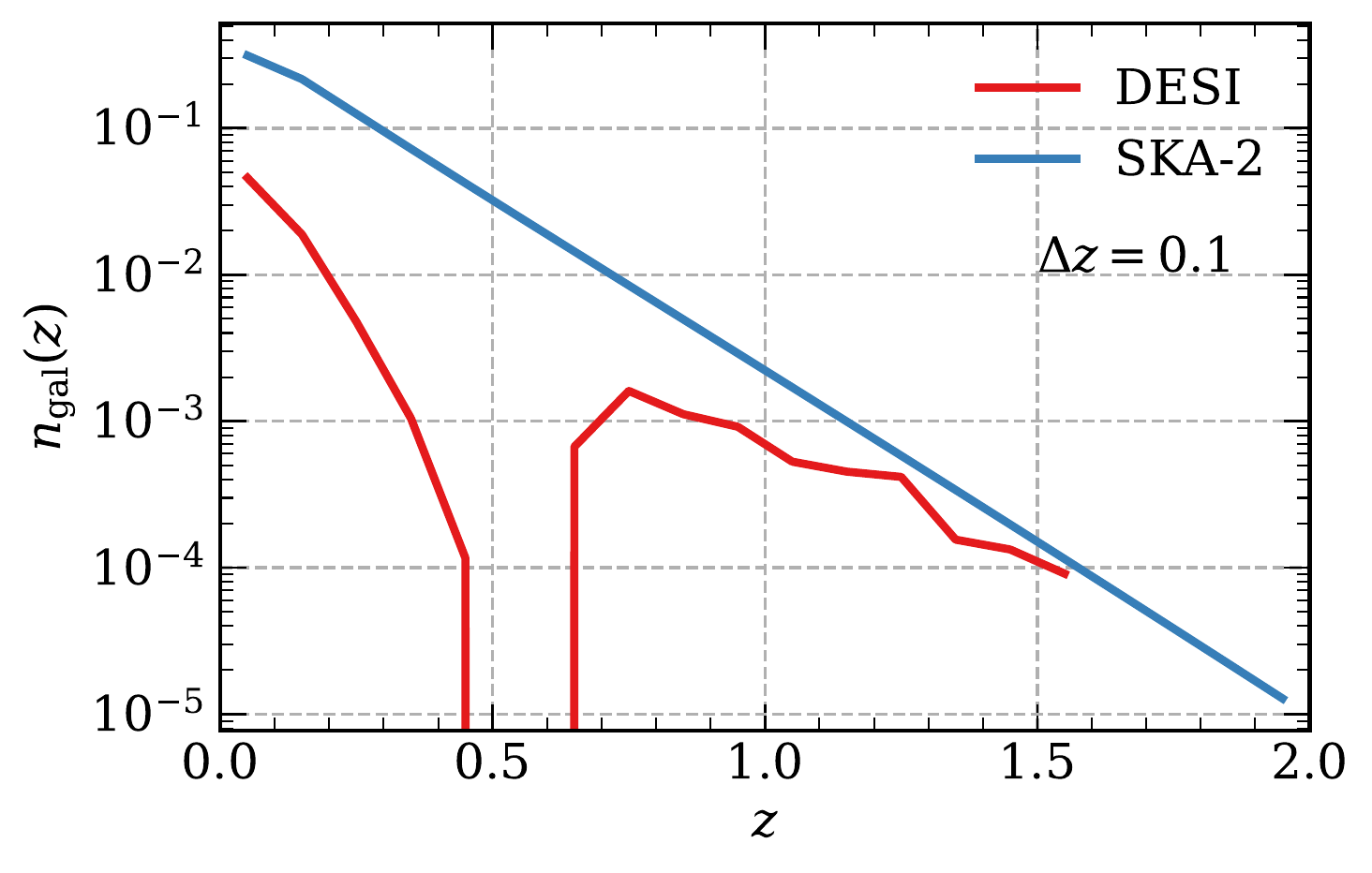}
    \caption{The expected redshift distribution of galaxy number densities $n_{\rm gal}(z)$ for DESI and SKA-2 with the redshift bin width $\Delta z=0.1$.}
    \label{fig:ngal_z}
\end{figure}
\begin{figure}
    \centering
    \includegraphics[width=\columnwidth]{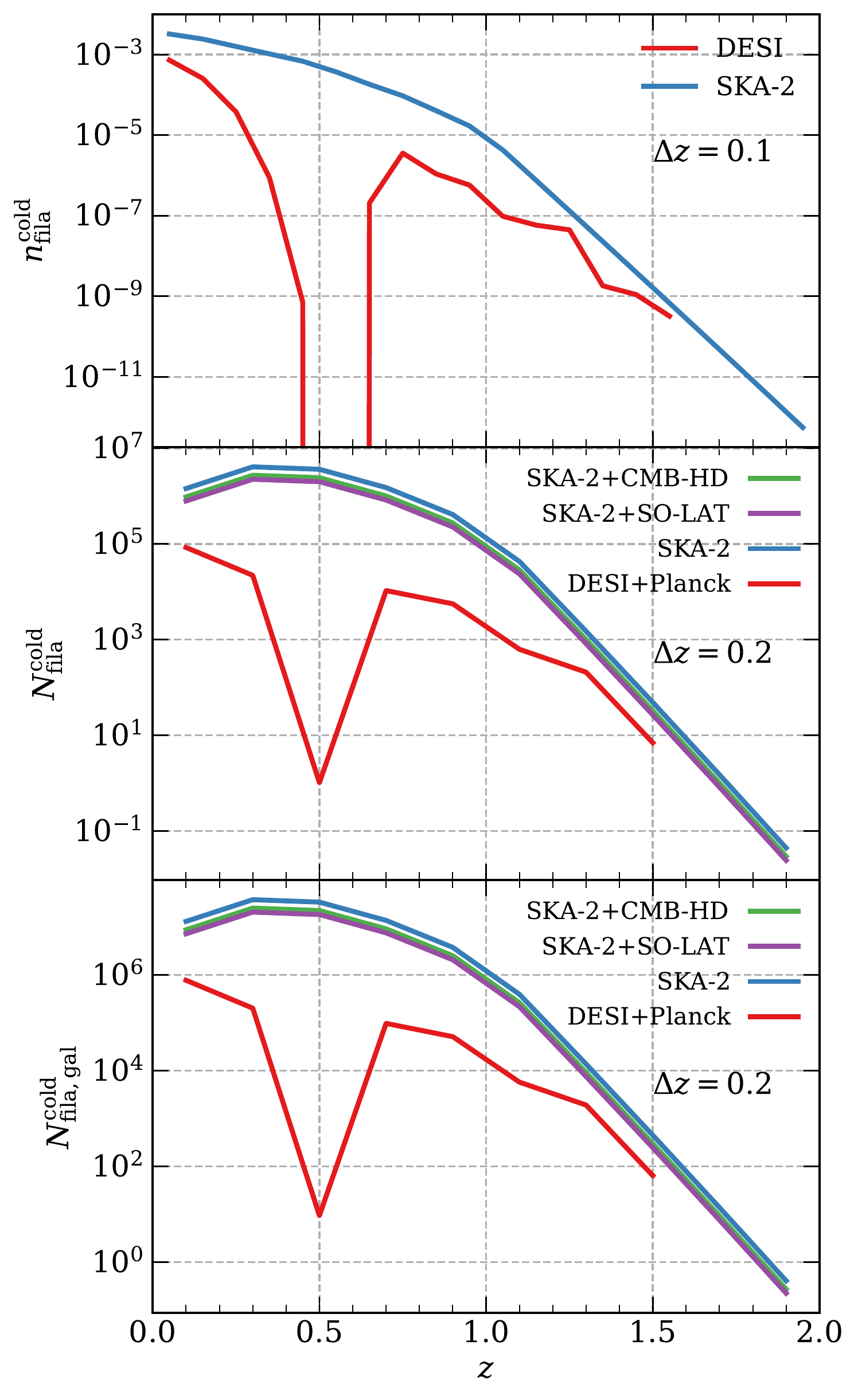}
    \caption{\textit{Top panel}: The expected number density of cold filaments $n_{\rm fila}^{\rm cold}(z)$ of DESI and SKA-2 with the redshift bin width $\Delta z = 0.1$. \textit{Middle panel}: The expected cold filament number $N_{\rm fila}^{\rm cold}$ with the redshift bin width $\Delta z = 0.2$.  \textit{Bottom panel}: The expected number of galaxies in cold filaments $N_{\rm fila,gal}^{\rm cold}$ with the redshift bin width $\Delta z = 0.2$.}
    \label{fig:nfila_Nfila_Ngal}
\end{figure}
For DESI, we use Table 2.3 of \citealt{DESI2016} to obtain the expected $n_{\rm gal}(z)$. We combine the $dN_{\rm ELG}/dz/d{\rm deg^2}$ and $dN_{\rm LRG}/dz/d{\rm deg^2}$ columns therein to get $dN_{\rm gal}/dz/d{\rm deg^2}$ at $z<1.6$, which is then transformed into the expected galaxy number density $n_{\rm gal}(z)$, as shown in Fig. \ref{fig:ngal_z}. Next we feed this DESI $n_{\rm gal}(z)$ into the $n_{\rm fila}^{\rm cold}-n_{\rm gal}$ relation to estimate the expected number density of cold filaments for DESI $n_{\rm fila}^{\rm cold}(z)$, and accounts for the difference of the survey volume to obtain the expected number $N_{\rm fila}^{\rm cold}(z)$ of cold filaments for DESI shown in Fig.~\ref{fig:nfila_Nfila_Ngal}. 

For the SKA-2 survey, we adopt the model for the $\rm HI$ galaxy number counts from semi-analytical simulations in \citealt{Yahya2015} to infer for the expected $n_{\rm gal}(z)$, focusing on the reference flux sensitivity. The galaxy number density model is 
\beq
\label{eq:ngal_SKA-2}
\frac{{\rm d}N(z)/{\rm d}z}{1 {\rm deg^2}} = 10^{c_1}z^{c_2}\exp(-c_3z)\,,
\eeq
in which the adopted best-fit parameters $c_1-c_3$ from semi-analytical simulations are shown in the 5th row of Table 4 in \citealt{Yahya2015}. Then we go through the same procedure as above to evaluate the resultant $n_{\rm fila}^{\rm cold}(z)$ and $N_{\rm fila}^{\rm cold}(z)$ of SKA-2 at $z<2$. They are presented in Fig.~\ref{fig:nfila_Nfila_Ngal}.

Finally, we adopt the number of galaxies per cold filament, 
\beq
\label{eq:gal_per_fila}
N_{\rm fila,gal}^{\rm cold,W21}/N_{\rm fila}^{\rm cold,W21} \approx 9.1\,, \no
\eeq
evaluated from the Wang21 catalog, and apply it to the $N_{\rm fila}^{\rm cold}(z)$ relation to estimate the number of filament galaxies $N_{\rm fila,gal}^{\rm cold}(z)$ for DESI and SKA-2, shown in the bottom panel of Fig.~\ref{fig:nfila_Nfila_Ngal}.


\bsp	
\label{lastpage}
\end{document}